\documentclass[12pt]{article}
\usepackage[margin=3cm]{geometry}
\usepackage{fourier}
\DeclareMathAlphabet{\mathcal}{OMS}{cmsy}{m}{n}
\usepackage{amsmath, amssymb, amsthm, mathrsfs,bm}
\usepackage{booktabs}
\usepackage{url}
\usepackage{graphicx}
\usepackage{enumitem}
\usepackage{algorithm}
\usepackage{algpseudocode}
\usepackage{comment}
\usepackage{xcolor}

\usepackage{graphicx}
\graphicspath{{image}}

\newcommand{\rom}[1]{\MakeUppercase{\romannumeral #1}} 

\algnewcommand{\Inputs}[1]{%
  \State \textbf{Inputs:}
  \Statex \hspace*{\algorithmicindent}\parbox[t]{.8\linewidth}{\raggedright #1}
}
\algnewcommand{\Initialize}[1]{%
  \State \textbf{Initialize:}
  \Statex \hspace*{\algorithmicindent}\parbox[t]{.8\linewidth}{\raggedright #1}
}

\algdef{SE}[SUBALG]{Indent}{EndIndent}{}{\algorithmicend\ }%
\algtext*{Indent}
\algtext*{EndIndent}

 \usepackage{setspace}
\let\Algorithm\algorithm
\renewcommand\algorithm[1][]{\Algorithm[#1]\setstretch{1.2}}

\usepackage[numbers]{natbib} 

\date{January 2024}

\begin{document}

\begin{titlepage}
    \centering
    \vspace*{\stretch{1}}
    {\LARGE Bayesian Inference for Stochastic Predictions of Non-Gaussian Systems with Applications in Climate Change\par}
    \vspace*{\stretch{1}}
    {\Large Yunjin Tong\par}
    \vspace{\stretch{0.5}}
    {\Large Undergraduate Mathematics Thesis\par}
    \vspace{\stretch{0.5}}
    {\large Advised by\par}
    \vspace{\stretch{0.1}}
    {\Large Professor Yoonsang Lee\par}
    \vspace*{\stretch{1}}
    \includegraphics[width=0.2\linewidth]{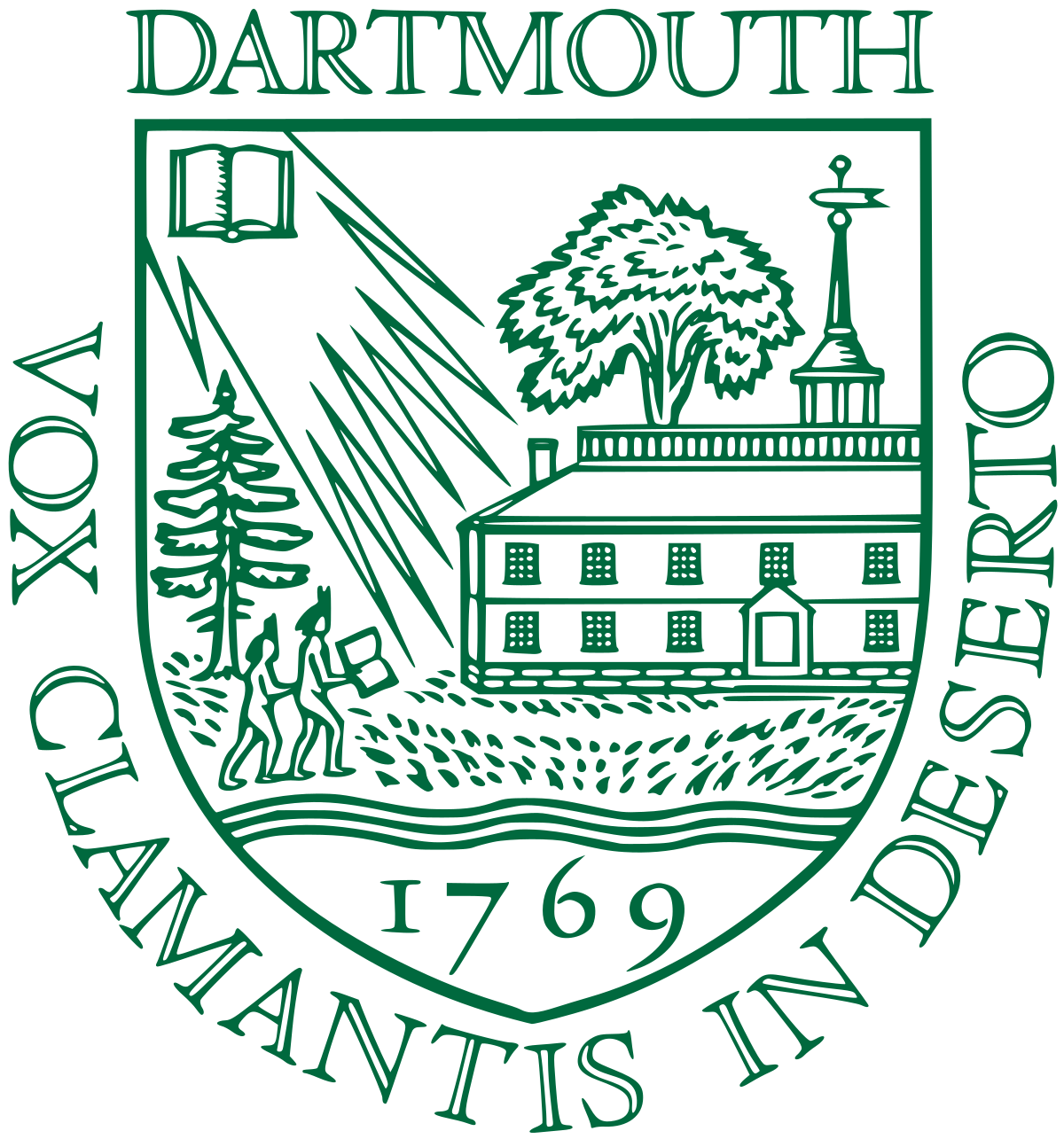}\par
    \vspace*{\stretch{0.1}}
    {\Large Dartmouth College\par}
    {\Large Hanover, New Hampshire\par}
    {\Large June, 2024\par}

\end{titlepage}

\section*{Abstract}
Climate change poses significant challenges for accurate climate modeling due to the complexity and variability of non-Gaussian climate systems. To address the complexities of non-Gaussian systems in climate modeling, this thesis proposes a Bayesian framework utilizing the Unscented Kalman Filter (UKF), Ensemble Kalman Filter (EnKF), and Unscented Particle Filter (UPF) for one-dimensional and two-dimensional stochastic climate models, evaluated with real-world temperature and sea level data. We study these methods under varying conditions, including measurement noise, sample sizes, and observed and hidden variables, to highlight their respective advantages and limitations. Our findings reveal that merely increasing data is insufficient for accurate predictions; instead, selecting appropriate methods is crucial. This research provides insights into issues related to information barrier, curse of dimensionality, prediction variability, and measurement noise quantification, thereby enhancing the application of these techniques in real-world climate scenarios.
\newpage
\section*{Acknowledgements}
Thanks to the generous support of the student grant from The Arthur L. Irving Institute and the Honors Thesis Grant from the Kaminsky Family Fund Award, administered by the Dartmouth Undergraduate Advising \& Research program.

I extend my heartfelt thanks to my supervisor, Professor Yoonsang Lee, for his unwavering support and profound inspiration. I am fortunate to have had the opportunity to take various undergraduate and graduate classes with him and to work on this thesis under his guidance. He has helped me
not only form a solid mathematical foundation but also gain invaluable life wisdom.

I also want to thank Professor Anne Gelb for her encouragement in pursuing a research career, along with the other professors and students who have taught and assisted me at Dartmouth.

Additionally, I acknowledge the critical data sources that supported this thesis. The global mean sea level are obtained using a weighted average of global tide gauge records. The data is from NOAA \citep{climate-gov}. The average of global surface temperature is from NASA \citep{nasa-climate}.

\newpage
\tableofcontents
\newpage
\section{Introduction}

Climate change is one of the most pressing global challenges facing humanity today. Accurate modeling of climate phenomena is crucial for understanding future scenarios and making informed policy decisions. However, the inherent complexity and multifaceted nature of climate systems present significant challenges for conventional modeling techniques. The deterministic models are often computationally expensive, requiring substantial computing resources to solve the complex equations governing fluid dynamics, thermodynamics, and other physical processes. Additionally, they face challenges due to sensitivity to initial conditions, the need to parameterize subgrid-scale processes, and biases resulting from model assumptions and simplifications.

In recent years, there has been growing interest in the role of stochastic processes in climate modeling. Stochastic models, which introduce elements of randomness and uncertainty, offer a more nuanced way of representing natural processes. These models are particularly useful for understanding events that are rare but consequential, like extreme weather conditions. However, integrating stochastic processes into climate models is not straightforward, especially when the systems under study do not conform to Gaussian distributions.

Moreover, Bayesian inference, a powerful statistical technique, provides a framework for incorporating uncertainty and prior knowledge into complex models. It offers a systematic approach for updating the probability of different outcomes based on new data, making it particularly well-suited for stochastic modeling. In the context of climate change modeling, one of the major advantages is that Bayesian inference can provide a probabilistic framework for predictions, meaning that it can offer not just point estimates but also credible intervals for parameters like temperature rise, precipitation changes, or sea level rise. This probabilistic aspect allows policymakers and scientists to assess risks more comprehensively. Moreover, Bayesian methods can continually incorporate new data to update these predictions, making the models more adaptive to new information. Therefore, using Bayesian inference in climate change models could lead to more targeted and effective climate policy, better allocation of resources for climate mitigation and adaptation, and ultimately, a more robust response to the uncertainties and risks posed by a changing climate.

To address the complexities of non-Gaussian systems in climate modeling, we propose a Bayesian framework that leverages filtering techniques: the Unscented Kalman Filter (UKF), Ensemble Kalman Filter (EnKF), and Unscented Particle Filter (UPF), which are applied independently to both one-dimensional and two-dimensional stochastic climate models with real-world temperature and sea level data. We evaluate the effectiveness and adaptability of these methods under different conditions, including varying measurement noise, sample sizes, as well as observed and hidden variables. This research aims to demonstrate their respective advantages and limitations, providing valuable insights for their application in real-world scenarios. Our research explores issues related to the information barrier, curse of dimensionality, variability of predictions, and quantification of measurement noise. We show that in data assimilation, simply obtaining more data is not sufficient; the selection of the appropriate methods and models is crucial for accurate and reliable predictions.

The thesis will be organized into several key sections: an introductory section and a related work section that outline the research background; a stochastic climate modeling section that elaborates on one-dimensional and two-dimensional models used to simulate climate change; a statistical inference section that details the mathematical foundations and algorithm implementations of three filtering methods; a parameter optimization and robustness testing section that explains the selection of key parameters; and two experimental sections, one for each model, presenting the findings. The paper concludes with a discussion of the implications of the results and potential avenues for future research.

\section{Related Work}
\subsection{Stochastic Modeling}
Stochastic modeling incorporates random processes and probabilistic approaches to simulate and predict complex systems where uncertainty and variability are inherent \citep{pinsky2010introduction}. In climate science, stochastic models are crucial for understanding and predicting changes in the climate system, which is influenced by a myriad of interconnected variables and inherently random natural processes.

Climate systems are highly complex and nonlinear, characterized by chaotic behaviors and underdetermined dynamics. Stochastic models help address these complexities by:
\begin{enumerate}
    \item Incorporating Uncertainties: They account for the random variations in climate variables such as temperature, pressure, and wind patterns \citep{palmer2005representing, palmer2009stochastic}. 
    \item Improving Ensemble Predictions: By generating multiple ensembles, they provide a range of possible outcomes that help in risk assessment and policy-making \citep{yiou2019stochastic,franzke2015stochastic}.
    \item Enhancing Model Accuracy: Stochastic elements can improve the realism of climate models by incorporating the probabilistic nature of various processes, from severe weather events to long-term climatic trends \citep{palmer2019stochastic,ewald2004accurate}.
\end{enumerate}

Stochastic modeling finds applications across various aspects of climate science. In weather forecasting, short-term climate models use stochastic methods to predict conditions by considering the probabilities of various meteorological inputs \citep{wilks1999weather}. For long-term climate projections, stochastic differential equations are utilized to model the uncertainty in future emissions and atmospheric reactions \citep{hasselmann1976stochastic,franzke2015stochastic,majda1999models,majda2001mathematical}. These models are also employed in paleoclimate reconstructions to infer historical climate conditions from proxy data, which are inherently noisy and uncertain \citep{tingley2012piecing}.

Despite their advantages, stochastic models face several challenges. Scalability issues arise when extending local models to global scales while managing computational loads. Parameter uncertainty also presents significant hurdles as determining accurate probabilities for various stochastic processes requires extensive data and sophisticated techniques. Moreover, integrating these models with traditional deterministic models to enhance predictiveness while maintaining realistic dynamics is complex \citep{franzke2015stochastic}.

Stochastic modeling is used in many climate models. One specific example is the adaptation of the Earth Balance Model \citep{budyko1969effect,sellers1969global,north1975theory,north1975analytical,north1981energy}, a simplified representation of the Earth’s climate system that calculates the balance between energy received from the Sun and the energy emitted back into space. The stochastic version of this model introduces random fluctuations in solar radiation, volcanic activity, and other climatic inputs to simulate their impacts on Earth’s energy balance and hence, on global temperature and climate variability \citep{imkeller2001energy}.

\subsection{Bayesian Filtering} 
Data assimilation is the process of integrating dynamical models with observational data to estimate the state of a dynamical system. This technique is fundamental in improving predictions and adjusting models based on new observations \citep{kalnay2003atmospheric,evensen2009data,majda2012filtering}. In particular, filtering based on Bayesian inference specifically addresses the challenge of estimating the state of a system as new observations become available in real-time. This is crucial across many fields, including science, engineering, and finance, where models often include complexities such as nonlinear dynamics and non-Gaussian distributions. These complexities make it difficult, if not impossible, to find analytical solutions, thus necessitating the use of sophisticated numerical algorithms for online filtering.

The original Kalman Filter (KF), developed by Rudolf E. Kalman in 1960, serves as the foundation for the filtering methods. Designed specifically for linear systems, it uses a set of mathematical equations to provide an efficient computational recursive solution based on the least-squares method \citep{kalman1960new}. The Extended Kalman Filter (EKF) emerged in the early 1960s as one of the first significant extensions of the Kalman Filter aimed at handling nonlinear systems. The practices and theoretical underpinnings of the EKF were notably elaborated in Anderson and Moore's work in 1979 \citep{anderson2012optimal}. The EKF adapts to nonlinearity by linearizing the state-space model's functions using Taylor series expansions and Jacobian matrices. This linearization occurs around the current estimate, adjusting the system's measurements and evolution models to maintain tractability. However, the approximation methods used in EKF can misrepresent actual nonlinear dynamics and the associated probability distributions, potentially leading to divergence where the filter's estimates deviate significantly from the true values.

To better address the complexities of nonlinear systems and avoid the pitfalls of EKF, EnKF and UKF are developed, each serving different specific needs and computational strategies. The EnKF, developed by Geir Evensen in 1994, utilizes an ensemble of forecasts to deal with nonlinear data assimilation problems \citep{evensen1994sequential,burgers1998analysis,houtekamer1998data}. This method explicitly acknowledges and handles errors in the data and model by maintaining a probability distribution (ensemble) that approximates the state’s probability distribution.
EnKF uses a Monte Carlo approach to represent the error covariance and state distribution through the use of an ensemble of forecasts. This method is particularly favored in large-scale geophysical models (like weather forecasting) where maintaining a full covariance matrix is computationally prohibitive.

The UKF was developed by Jeffrey Uhlmann and Simon Julier in 1997 based on the intuition that it is easier to approximate a Gaussian distribution than it is to approximate arbitrary nonlinear functions \citep{julier1997new}. It addresses the shortcomings of the EKF by using a deterministic sampling technique known as the Unscented Transform to pick a minimal set of sample points (sigma points) around the mean. These points are propagated through the nonlinear functions, from which a new mean and covariance estimates are calculated. UKF utilizes a set of deterministically chosen sample points to capture the true mean and covariance of the prior distribution more accurately than the EKF. It is highly effective in smaller-dimensional systems where the computational overhead of using an ensemble is not necessary. The UKF has, however, the limitation that it does not apply to general non-Gaussian distributions.

Another popular solution strategy for the general filtering problem is to use sequential Monte Carlo methods, also known as particle filters (PFs) \citep{gordon1993novel, doucet2001sequential}. These methods provide a complete representation of the posterior distribution of the states, allowing for the computation of statistical estimates such as the mean, modes, kurtosis, and variance. Thus, they are capable of handling any nonlinearities or non-Gaussian distributions effectively. Particle filtering captures non-Gaussian features by assigning different weights to different samples (or particles). This technique is particularly effective for low-dimensional dynamical systems and does not require any assumptions about the prior distribution, leading to consistent Bayesian posterior statistics \citep{doucet2001sequential, van2009particle}. The particle weights also play a crucial role in determining which particles to keep during resampling strategies \citep{li2015resampling}, which helps to prevent particle degeneracy—a common problem where only a small number of particles end up with significant weights while the rest are virtually ignored. However, despite its effectiveness in low-dimensional settings, particle filtering is limited by the curse of dimensionality in high-dimensional systems, necessitating exponentially more particles as the dimensionality increases \citep{bengtsson2008curse, snyder2008obstacles}. Moreover, PFs rely heavily on importance sampling to function correctly, which requires carefully designed proposal distributions that can approximate the posterior distribution well. Designing these proposals is challenging, especially when new measurements are outliers or the likelihood is extremely peaked relative to the prior. These scenarios are particularly common in fields such as engineering and finance, where sensors may be highly accurate or data may exhibit sudden, non-stationary shifts \citep{pitt1999filtering}. To address these issues, various linearization techniques have been developed, such as using the EKF Gaussian approximation as the proposal distribution within a particle filter \citep{de2000sequential, de2003bayesian, pitt1999filtering, doucet1998sequential}.

Building on these concepts, the UPF was proposed by Merwe et al., utilizing the Gaussian approximation methods of the UKF as the proposal distribution within a particle filter framework, effectively merging the strengths of both the UKF and particle filters \citep{van2000unscented}. This approach is particularly advantageous when dealing with highly nonlinear processes or models that include heavy-tailed noise. The UKF is adept at producing proposal distributions that have a larger support overlap with the true posterior compared to those generated by the EKF, enhancing the efficacy of the sampling process. Moreover, the UKF's ability to theoretically model heavier tails than the EKF, while integrating the latest observational data before the evaluation of importance weights, allows the UPF to perform exceptionally well in scenarios with peaked likelihoods or when data contains outliers. The UPF retains the particle filter’s ability to capture non-Gaussian features effectively, but it also inherits the curse of dimensionality, which can be a significant limitation in high-dimensional systems. This characteristic necessitates a larger number of particles to maintain accurate state estimations, echoing the inherent challenges faced by traditional PFs.

\subsection{Applications of Bayesian Filtering} 

Filtering techniques are widely applied across various disciplines including science, engineering, and finance. Each field leverages the distinct features of filters to serve specialized purposes, such as improving accuracy in state estimation, enhancing noise reduction, and facilitating robust data assimilation in complex dynamic systems. These applications utilize the unique properties of each filter type to address the specific challenges and requirements inherent in diverse environments. In engineering, filtering techniques are widely applied to problems in aerospace \citep{grewal2010applications}, automotive tracking \citep{coue2006bayesian,floudas2005survey}, materials science \citep{gyorgy2014unscented}, singal processing \citep{roth2017ensemble} and remote sensing \citep{chatzi2009unscented}, among others. These methods address the complexities and nonlinearities inherent in these fields.

EnKF is commonly associated with geosciences \citep{evensen1996assimilation,sebacher2013probabilistic}, and also extensively used in hydrology, environmental modeling \citep{szunyogh2005assessing}, and reservoir engineering \citep{aanonsen2009ensemble}, where the challenges of modeling large spatial systems and assimilating massive data sets are prevalent. EnKF has shown particular efficacy in atmospheric and oceanic data assimilation \citep{houtekamer2005atmospheric,whitaker2008ensemble,szunyogh2008local,houtekamer2009model}.

PFs have been widely adopted in robotics, a field characterized by non-Gaussian and highly nonlinear environments. Their robustness and adaptability make them ideal for dynamic and uncertain scenarios commonly encountered in robotic navigation and mapping \citep{thrun2002particle,fox2001particle}. Moreover, PFs are especially beneficial in finance, where markets frequently exhibit non-Gaussian behaviors, particularly in the tails of distributions, such as heavy tails in asset returns. PFs are adept at adaptively approximating any posterior distribution without assuming a specific form for the error distribution, providing crucial flexibility for accurately modeling and predicting financial risks and outcomes \citep{Kearns2005,javaheri2003filtering,wells2013kalman,date2011linear}.

UPF has been modified and utilized extensively for tracking applications, demonstrating significant improvements in performance and reliability \citep{zhan2008modified,havangi2018target}. Recently, UPF has been applied to critical tasks in energy storage, such as estimating the life and discharge times of lithium-ion batteries, further underscoring its utility in practical and high-stakes applications \citep{zhang2018improved, miao2013remaining,wang2020framework}.

\section{Stochastic Climate Modeling}

\subsection{One-Dimensional Model}
We use a Earth's Planetary Balance model as the basis for the one-dimensional ordinary differential equation (ODE) model. The use of the model is inspired by MIT's course \citep{MITClimateModelFall2023}. The conceptual idea of the Earth's Planetary Balance model is as below:

\begin{align}
&  \text{ Change in Heat Content} = \notag \\
& + \text{{Absorbed Solar Radiation (energy from the Sun's rays)}}\notag \\
& - \text{{Outgoing Thermal Radiation (i.e. blackbody cooling to space)}} \notag \\
& + \text{{Human-Caused Greenhouse Effect (trapped outgoing radiation),}}
\end{align}

where each of these terms is interpreted as an average over the entire globe \citep{MITClimateModelFall2023,north1981energy}.

\begin{figure}[h!]
\centering
\includegraphics[width=0.8\linewidth]{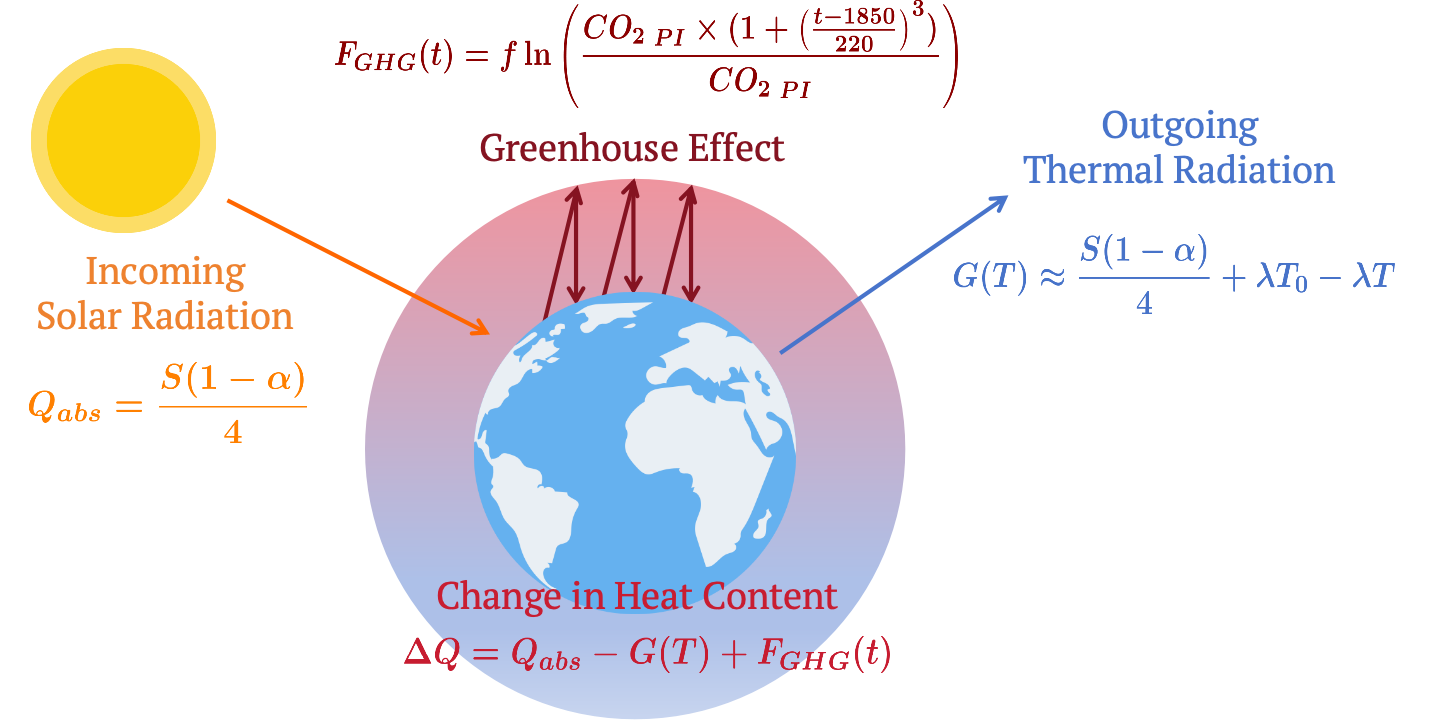}
\caption{Earth's Planetary Balance Model}

\label{fig:EPB}
\end{figure}

For simplicity, we use the following formula to represent the above Earth's Planetary Balance model:
\begin{equation}
\Delta Q = Q_{abs} - G(T)+ F_{GHG}(t),
\label{model}
\end{equation}
where $\Delta Q$ represents the change in heat content, $Q_{abs}$ represents the absorbed solar radiation, $G(T)$ represents the outgoing thermal radiation as a function of temperature, which is subtracted since it's energy leaving the Earth. $F_{GHG}(t)$ is the forcing due to greenhouse gases, which is added since it represents additional heat retained by the Earth's system due to human activities. An illustration of the Earth's Planetary Balance Model is provided in Figure~\ref{fig:EPB}, offering a visual representation of these dynamics.

\subsubsection{Absorbed Solar Radiation}
At the distance Earth orbits the Sun, the intensity of sunlight reaching Earth is measured as solar insolation, which is \(S = 1368 W/m^2\). The Earth reflects a portion of this sunlight back into space, a process quantified by a small fraction albedo, or planetary reflectivity \(\alpha = 0.3\). This reflectivity accounts for the light bounced off surfaces such as white clouds, snow, and ice. Consequently, the fraction of sunlight absorbed by the Earth is \( (1 - \alpha) \). Given that sunlight arrives on Earth as nearly parallel rays, the Earth's cross-section intercepting these rays can be represented as a disk with an area of \( \pi R^2 \). However, since the calculations for other terms in model~\eqref{model} apply to the Earth’s entire surface area, which is \( 4\pi R^2 \), the per unit surface area of absorbed solar radiation needs to be adjusted by a factor of 4. Therefore, we can model the absorbed solar radiation as below \citep{MITClimateModelFall2023}:

\begin{equation}
  Q_{abs} = \frac{S(1 - \alpha)}{4}. \label{abs}
\end{equation}

\subsubsection{Thermal Radiation}

The term $G(T)$ in the model denotes the net effect of both negative feedback mechanisms, like blackbody radiation, that dampen warming, and positive feedbacks, such as the water vapor feedback, that enhance warming. Given the complexity of these interactions, the model simplifies them by only considering the first term of a Taylor Series expansion around the pre-industrial average temperature $T_0 = 14.0^\circ C$ \citep{MITClimateModelFall2023}:

\begin{equation}
G(T) \approx G(T_0) + G'(T_0)(T - T_0) = G'(T_0)T + (G(T_0) - G'(T_0)(T_0)).
\end{equation}

To simplify the expression, we define:
\begin{equation}
A = G(T_0) - G'(T_0)(T_0), \quad \lambda = -G'(T_0) \quad 
\end{equation}
which gives
\begin{equation}
G(T) \approx A - \lambda T \label{thermal}.
\end{equation}

The minus sign in front of $T$ indicating it restores equilibrium.
$\lambda$ is the climate feedback parameter. The value that is chosen for $\lambda$ is from the review paper \citep{sherwood2020assessment}, which produced the most comprehensive and up-to-date estimate of the climate feedback parameter: $\lambda \sim \mathcal{N}(-1.3, 0.4)$.
Thus, we use $\lambda = -1.3  W/m^2/^\circ C $  as the climate feedback parameter in this thesis.

The value of parameter $A$ is determined by recognizing the preindustrial equilibrium state, where Earth's energy budget was in a perfect balance prior to human influence: 
\begin{equation}
Q_{abs} = G(T_0)
\end{equation}
or
\begin{equation}
\frac{S(1 - \alpha)}{4} = A -  \lambda T_0.
\end{equation}

By rearranging this equation, we find that the value of $A$ in $W/m^2$ is given by
\begin{equation}
A = \frac{S(1 - \alpha)}{4} +  \lambda T_0. \label{A}
\end{equation} 

\subsubsection{Human-caused Greenhouse Effect}
The greenhouse effect is empirically understood to correlate logarithmically with the concentrations of carbon dioxide CO$_2$ in the atmosphere. The human-caused greenhouse effect is modeled as:
\begin{equation}
 F_{GHG} = f \ln \left( \frac{CO_2(t)}{{CO_2}_{\;PI}} \right), \label{ghg}
\end{equation}
where $f = 5.0 W/m^2$ is the forcing coefficient and ${CO_2}_{\;PI} = 280.0 ppm$ is is the pre-industrial $CO_2$ concentration. \( CO_2(t) \) is a function that gives the $CO_2$ concentration at time \( t \).

Human greenhouse gas emissions have significantly changed the Earth's energy equilibrium, deviating from the stable climatic conditions that prevailed for millennia before the industrial era. Considering that human CO$_2$ emissions are a primary contributor to  global warming, our model anticipates that inputting historical CO$_2$ concentration increases should approximate the observed rise in global temperatures. Starting from the onset of the industrial revolution in 1850, the increase in CO$_2$ levels can be fairly accurately captured by the following cubic equation \citep{MITClimateModelFall2023}:

\begin{equation}
CO_2(t) = {CO_2}_{\;PI} \times (1 + \left(\frac{t - 1850}{220}\right)^3).
\end{equation}

\subsubsection{Change in Heat Content}

The heat content, denoted as $CT$, is determined by the temperature $T$ measured in Kelvin and the heat capacity of the Earth's climate system. Consequently, the change in heat content over time can be simply represented by $\frac{d(CT)}{dt}$.  Although our primary focus is on the atmospheric temperature, which possesses a relatively small heat capacity, its heat is closely coupled with that of the upper ocean, which exhibit a significantly larger heat capacity of $C = 51 J/m^2/^\circ C$.
Given that the heat capacity of ocean water remains relatively constant across temperatures, this can be reformulated in terms of the change in temperature with time as \citep{MITClimateModelFall2023}:

\begin{equation}
\Delta Q  = C \frac{dT}{dt}
\end{equation}

Combining all of these subcomponent models, namely the absorbed solar radiation model \eqref{abs}, the outgoing thermal radiation model \eqref{thermal}, and the human-caused greenhouse effect model \eqref{ghg}, we formulate the governing equation of the mode;~\eqref{model} as an ODE:
\begin{equation}
C \frac{dT}{dt} = \frac{(1 - \alpha)S}{4} - (A -  \lambda T) + f \ln \left(\frac{CO_2(t)}{{CO_2}_{\;PI}}\right), \label{contimodel}
\end{equation}
which serves as the governing equation for the time evolution of Earth's globally-averaged surface temperature. Table \ref{tab:EBM} summarizes the parameters we use in the model's setup.

\begin{table}
  \caption{Setup of Earth’s Planetary Balance model.}
  \centering
  \setlength{\tabcolsep}{3mm}{
  \begin{tabular}{p{2.5mm} cc}
  \toprule
  Parameter & Value &  Description \\
  \midrule
  $f$& 5 &\(CO_2\) forcing coefficient \([W/m^2]\) \\
  $CO_{2\;PI}$ & 280 & Pre-industrial \(CO_2\) concentration \([ppm]\) \\
  $\lambda$ & -1.3 & Climate feedback parameter \([W/m^2/^\circ C]\)\\
  $C $ & 51 & Atmosphere and upper-ocean heat capacity \([J/m^2/^\circ C]\)\\

  $\alpha$ &0.3& Planet albedo, \(0.0–1.0\) [unitless] \\
  S &1368& Solar insolation \([W/m^2]\) \\
  A & $S\times (1 - \alpha)/4 + \lambda\times T_0$ & Linearized outgoing thermal radiation: offset \([W/m^2]\) \\
   
  \bottomrule
  \end{tabular}}
  \label{tab:EBM}
\end{table}

\subsubsection{Numerical Method and Stochastic Process}
We employ the Euler method to discretize the exact ODE above \eqref{contimodel} in time. By truncating the Taylor series expansion to the first order, we approximate the ODE \eqref{contimodel} as 
$\Delta t \rightarrow 0$.
\begin{equation}
C\frac{T(t + \Delta t) - T(t)}{\Delta t} = \frac{(1 - \alpha)S}{4} - (A -  \lambda T(t)) + f \ln \left(\frac{CO_2(t)}{{CO_2}_{\;PI}}\right).
\label{discre}
\end{equation}

Next, we use the subscript $n$ to denote the $n$-th timestep, where $T_{n+1} = T(t_{n+1})$ denotes the temperature at the next timestep $t_{n+1} = t_n + \Delta t$.

By rearranging the equation~\eqref{discre}, we can solve for the temperature at the next timestep $n + 1$ based on the temperature at the present timestep $n$:
\begin{equation}
T_{n+1} = T_n + \Delta t \times \frac{1}{C} \left[\frac{(1 - \alpha)S}{4} - (A - \lambda T_n) + f \ln\left(\frac{CO_{2,n}}{{CO_2}_{\;PI}}\right)\right].\label{nume}
\end{equation}

Recall equation \eqref{A}, substitute $A$ in \eqref{nume}, we get a simplified version:
\begin{equation}
T_{n+1} = T_n + \Delta t \times \frac{1}{C} \left[  \lambda (T_n-T_0) + a \ln\left(\frac{CO_{2,n}}{{CO_2}_{\;PI}}\right)\right]. \label{ODE}
\end{equation}

To better reflects the inherent uncertainties and variabilities in the Earth's climate system, we introduces an element of stochasticity by incorporating random fluctuations into the model. This gives us the Stochastic Ordinary Differential Equation (SODE) model as below:

\begin{equation}
T_{n+1} = T_n + \Delta t \times \frac{1}{C} \left[  \lambda (T_n-T_0) + f \ln\left(\frac{CO_{2,n}}{{CO_2}_{\;PI}}\right)\right] + \sigma \Delta W_{n}.\label{discSDE}
\end{equation}

Here, \( \Delta W_n \) represents the increment of the Wiener process during the time step from \( t_n \) to \( t_{n+1} \), which is normally distributed with mean 0 and the standard deviation  $\sigma$. This term is the discrete equivalent of the continuous-time Gaussian white noise term \( dW_t \) in the SODE.

The temperature data we use in our one-dimensional model is from NASA \citep{nasa-climate}, which provides temperature anomaly data. We add a constant, which is the long-term average from 1951 to 1980, to the anomaly data to obtain the actual temperature data.

\subsubsection{Statistical Properties}
\begin{figure}[h!]
\centering
\includegraphics[width=0.6\linewidth]{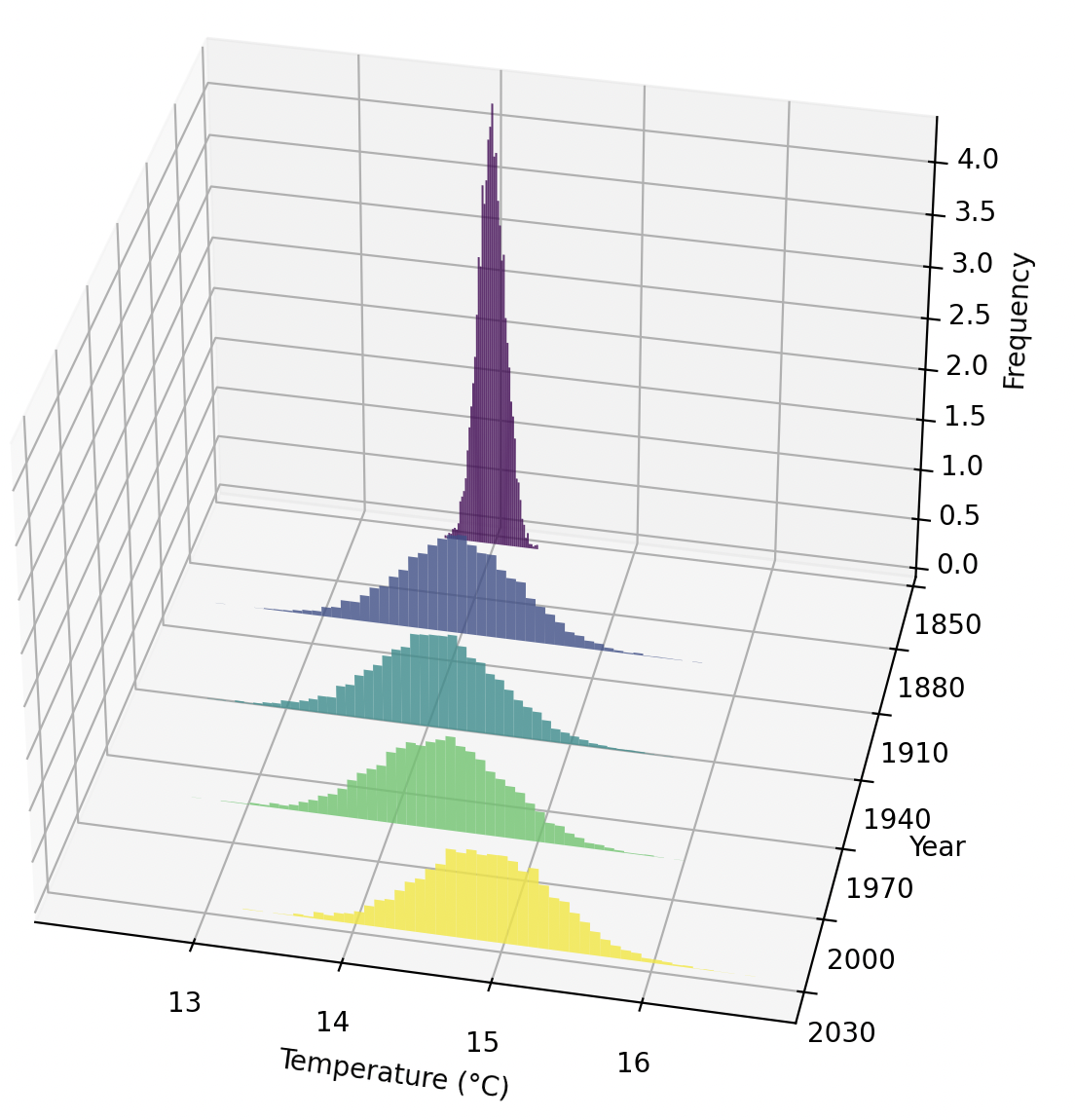}
\caption{Evolution of the temperature distribution at 43-year intervals from 1850.}

\label{fig:temphist}
\end{figure}

\begin{figure}[h!]
\centering
\includegraphics[width=0.6\linewidth]{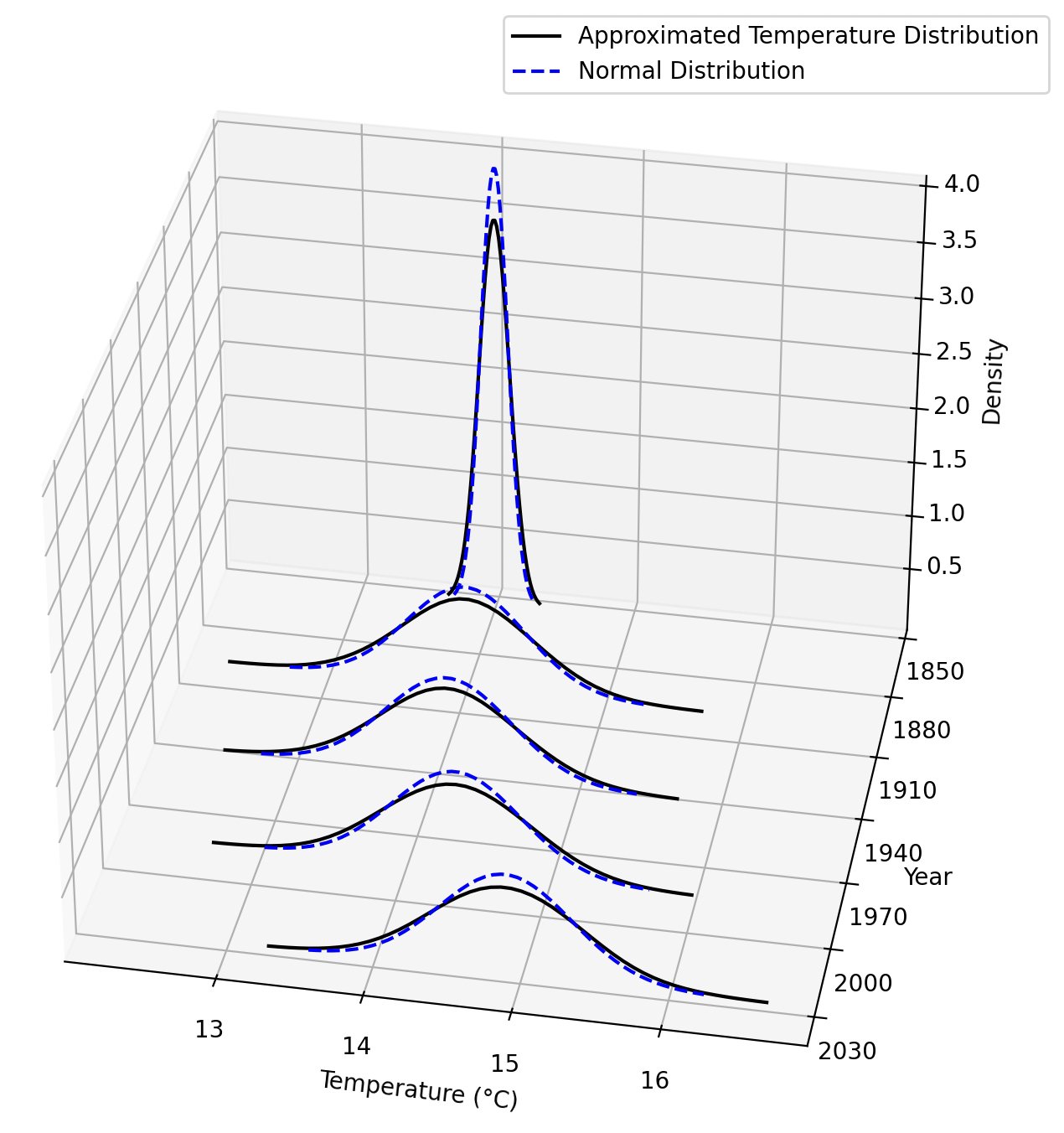}
\caption{Comparison of the approximated temperature distribution against a normal distribution with matching mean and standard deviation, highlighting deviations from normality. The normal distribution is depicted with a three-standard-deviation range.}

\label{fig:tempdis}
\end{figure}

We conducted 1000 simulations comprising 144 time steps each from 1850 to 2023, and the resulting temperature distributions are depicted as histograms at every 43 years in Figure~\ref{fig:temphist}. Additionally, we plot the approximated probability density function of the temperature distribution alongside a normal distribution with identical mean and standard deviation in Figure~\ref{fig:tempdis}.

Furthermore, we compute the discrete Kullback-Leibler (KL) divergence from the temperature distribution to the normal distribution, employing 1000 samples. Denote by \(X\) the set of all possible events, \(P\) the temperature distribution, and \(Q\) the normal distribution. The probability mass functions for a discrete random variable \(x\) under distributions \(P\) and \(Q\) are \(P(x)\) and \(Q(x)\), respectively. The KL divergence from \(Q\) to \(P\) is defined as:
\begin{equation}
D_{KL}(P \parallel Q) = \sum_{x \in X} P(x) \log\left(\frac{P(x)}{Q(x)}\right).
\end{equation}
Table~\ref{tab:KL} presents the KL divergence values measured at 43-year intervals. A discrete KL divergence is also calculated between two normal distributions as a reference, yielding a value of 0.015.

\begin{table}[h!]
  \caption{KL Divergence values from the temperature distribution to the normal distribution at 5 different years with 43-year intervals.}
  \centering
  \setlength{\tabcolsep}{5mm}{
  \begin{tabular}{c c}
  \toprule
Year & KL Divergence \\
  \midrule
1850 & 0.034 \\
1893 & 0.052 \\
1936 & 0.045 \\
1979 & 0.055 \\
2021 & 0.047 \\
  \bottomrule
  \end{tabular}}
  \label{tab:KL}
\end{table}

Observations from Figures~\ref{fig:temphist} and~\ref{fig:tempdis} indicate that there are both left and right tails, signifying a rise in extreme  temperature events. The flattening of the approximated temperature distribution suggests relatively low kurtosis, indicating a less pronounced concentration of temperatures around the mean, which contrasts with the more peaked spread of a normal distribution. The actual temperature distributions' deviation from the blue lines of the normal distribution underscores the potential inadequacy of simple Gaussian models in capturing the subtleties of temperature changes in a climate impacted by global warming. Furthermore, the KL divergence values listed in Table~\ref{tab:KL} consistently exceed the reference KL divergence value calculated between two normal distributions, which also indicates the non-normal distribution pattern in temperature changes.
In summary, the temperature distribution, although not markedly different from a normal distribution, is distinctly non-normal, primarily due to the rising incidence of extreme temperature events.

\subsection{Two-Dimensional Model}

Both the sea level the global temperature have fluctuated significantly throughout history. Three million years ago, they were both higher than current levels by 25 to 35 meters and 2-3$^\circ$C, respectively \citep{rahmstorf2007}. However, just 20,000 years ago the sea level was 140 meters lower and the temperature was 4-7$^\circ$C cooler than current numbers \citep{church2011sea}. The impact of global warming on sea-level rise and its hydrologic and environmental consequences has attracted worldwide attention because of its profound implications for coastal communities, ecosystems, and economies. Scientists and mathematicians have created models and estimates to predict sea level and temperature rises, either separately or in conjunction with each other. The inherent relationship between temperature and sea-level rise is embedded in the historic data. Made on the basis of the analysis of historic data from 1880 to 2001, Rahmstorf developed a semiempirical approach to project the sea-level rise in 2007. In this approach, the rate of sea-level rise is found to be proportional to the temperature with a constant of 3.4 mm/year per °C, and the sea level in 2100 was predicted to rise within the range of 0.5 to 1.4 m above the 1990 level \citep{rahmstorf2007}. However, these predictions are considerably greater than
those published in the Third Assessment Report of the Intergovernmental Panel on Climate Change (IPCC) \citep{church2001changes}. There were a few other issues with Rahmstorf’s model that contributed to the discrepancies. First, the historical temperature data were averaged over five-year periods, reducing the data set from 122 to 24 pairs and potentially distorting the original time-series relationship. Second, sea-level rise was assumed to be directly proportional to temperature change, with temperature as a fixed input. The model overlooked the sea level's temporal state, which can act as a negative feedback and slow the rise based on hydrologic and atmospheric principles. Additionally, while the model considered the effect of temperature on sea-level rise, it did not account for the reverse influence—how sea-level rise could affect temperature through increased water vapor due to higher temperatures and greater sea surface area.

Based on Rahmstorf’s model, Aral et al. introduced a system of coupled differential equations to describe the dynamic changes in temperature and sea level over time \citep{aral2012dynamic}. They solved this system of linear differential equations and found a 90\% confidence interval around the predicted temperature and sea level rise.
This approach addressed a significant limitation in Rahmstorf’s study by accounting for the interdependency of the two variables the two variables. Furthermore, the model used two year averaging to smooth out oscillation without losing as much precision as five year averaging. They applied their models to forecast outcomes from 1950 to 2001, comparing projected values with actual data. Their sea level model demonstrated high accuracy, while the temperature model, while close, was slightly less precise. Extending their projections to 2100, they estimated a 1.33°C temperature increase and a 423.77mm sea level rise, consistent with IPCC estimates \citep{aral2012dynamic}.

The model developed by Aral et al. is based on the interaction between global temperature change and sea-level rise, more specifically the proportional relationship between the rate of sea level rise and temperature. This linear relationship may be extended to approximately describe the behavior of the complex dynamic system by the governing coupled ODE:

\begin{equation}
\left\{
\begin{aligned}
\frac{dT(t)}{dt} &= a_{11}T(t) + a_{12}H(t) + c_{1} \\
\frac{dH(t)}{dt} &= a_{21}T(t) + a_{22}H(t) + c_{2}
\end{aligned}
\label{2dode}
\right.
\end{equation}
where $t$ = time; $T(t)$ = global mean surface temperature at time $t$; $H(t)$ = global sea level at time $t$; $\frac{dT(t)}{dt}$ and $\frac{dH(t)}{dt}$ = rate of temperature change and sea-level rise, respectively; $a_{ij}$ and $c_i(i,j = 1, 2)$ = constant coefficients. Coefficients $a_{ij}(i \neq j)$ reflect the effect of temperature change and sea-level rise on each other, whereas coefficients $a_{ii}(i = j)$ represent the temporal feedback of the state variables on themselves \citep{aral2012dynamic}. 

Aral et al. provide a set of coefficients for their model, which we have slightly adjusted to develop a set of coefficients for our implementation, shown in Table \ref{table:2dcoefficients}.

\begin{table}[h!]
\centering
\begin{tabular}{ccc}
\toprule
Coefficient  & Value & Description\\
 \midrule
\( a_{11} \) & -0.16 & Temporal Feedback of Temperature \\
\( a_{12} \)  & 0.008 & Coupling Coefficients \\
\( c_{1} \)    & 0.0187 & Baseline Rate of Temperature\\
\( a_{21} \) & 0.4673 & Coupling Coefficients\\
\( a_{22} \)   & -0.0145 & Temporal Feedback of Sea Level\\
\( c_{2} \)    & 0.2072 & Baseline Rate of Sea Level\\
\bottomrule
\end{tabular}
\caption{Adjusted coefficients for our two-dimensional model.}
\label{table:2dcoefficients}
\end{table}

Based on \eqref{2dode}, we develop a two-dimensional SODE model through discretizing \eqref{2dode} in time and adding noise terms.

\begin{equation}
\left\{
\begin{aligned}
T_{n+1} &= T_n + (a_{11}T_n + a_{12}H_n + c_1) \Delta t + \sigma_1 \Delta W_{1,n}, \\
H_{n+1} &= H_n + (a_{21}T_n + a_{22}H_n + c_2) \Delta t + \sigma_2  \Delta W_{2,n}.
\end{aligned}
\right.
\label{2dmodel}
\end{equation}

\begin{figure}[!htbp]
\centering
\includegraphics[width=0.5\linewidth]{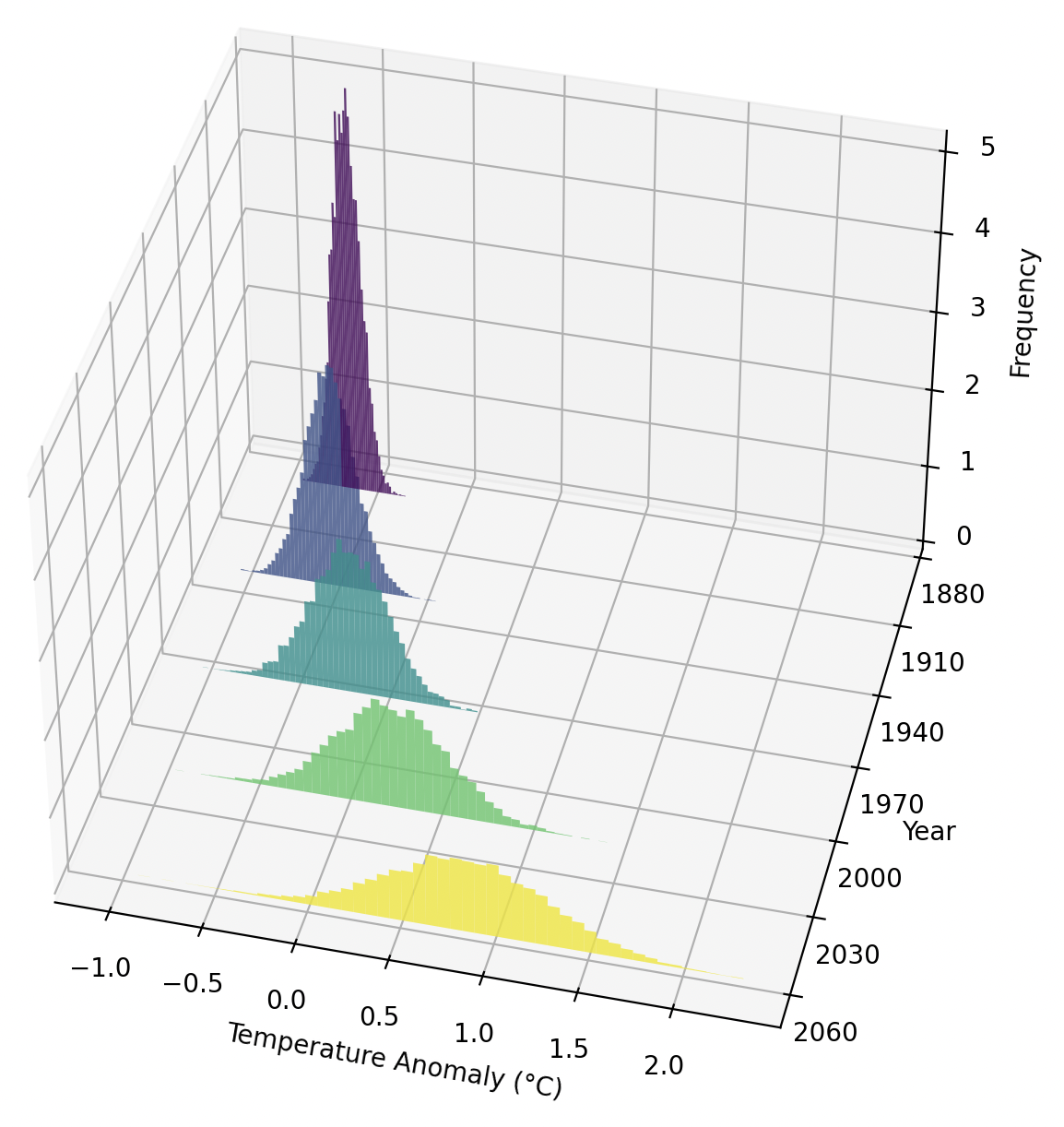}
\caption{Evolution of the temperature anomaly distribution at 43-year intervals from 1880.}

\label{fig:temphist2}
\end{figure}

\begin{figure}[!htbp]
\centering
\includegraphics[width=0.5\linewidth]{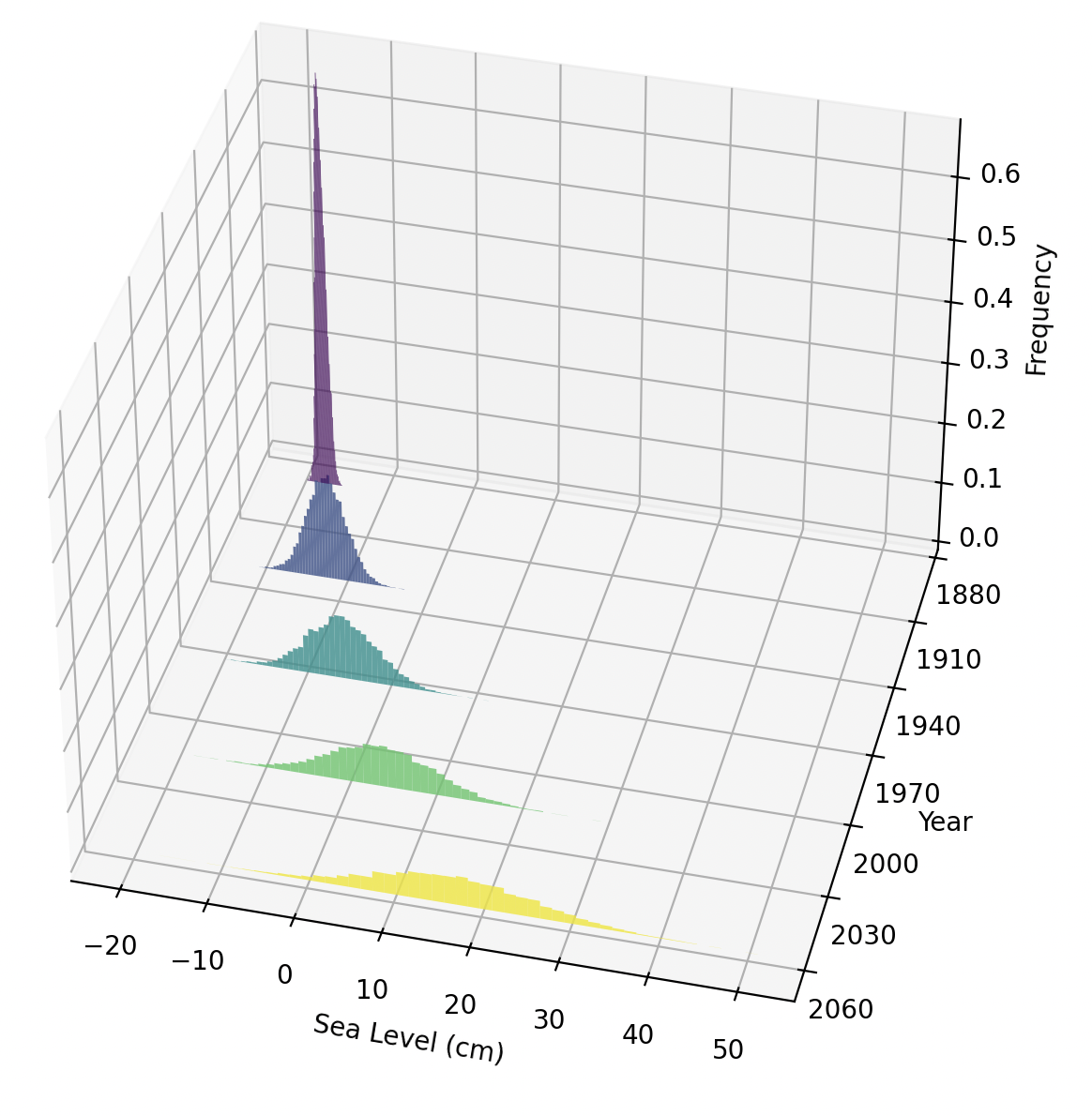}
\caption{Evolution of the sea level distribution at 43-year intervals from 1880.}

\label{fig:seahist}
\end{figure}

Here, \( T_n \) and \( H_n \) represent the discrete values of temperature and sea level at time step \( n \), respectively. \( \Delta t \) signifies the time step size, while \( \Delta W_{1,n} \) and \( \Delta W_{2,n} \) stand for independent Gaussian random variables with mean zero and standard deviation one, representing the Wiener increments. Additionally, \( \sigma_1 \) and \( \sigma_2 \) denote the noise amplitudes for temperature and sea level, respectively.

Since this model is developed using temperature anomaly data and sea-level data in cm, we use temperature anomaly data directly from NASA \citep{nasa-climate} and sea-level data from NOAA \citep{climate-gov}, dividing the latter by a factor of 10.

\subsubsection{Statistical Properties}

\begin{figure}[h]
\centering
\includegraphics[width=0.5\linewidth]{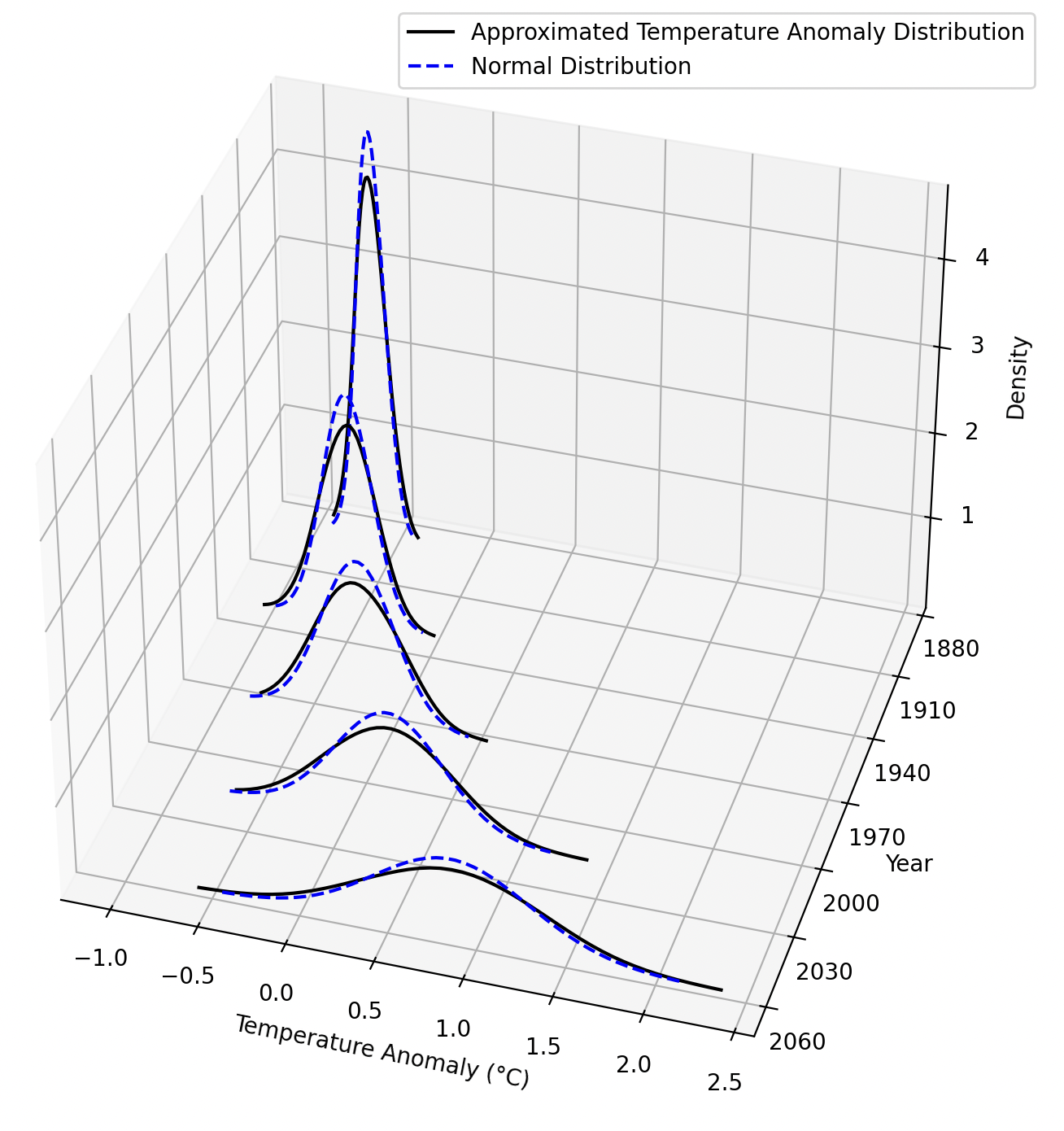}
\caption{Comparison of the approximated temperature anomaly distribution against a normal distribution with matching mean and standard deviation. The normal distribution is depicted with a three-standard-deviation range.}

\label{fig:tempdis2}
\end{figure}

\begin{figure}[h]
\centering
\includegraphics[width=0.5\linewidth]{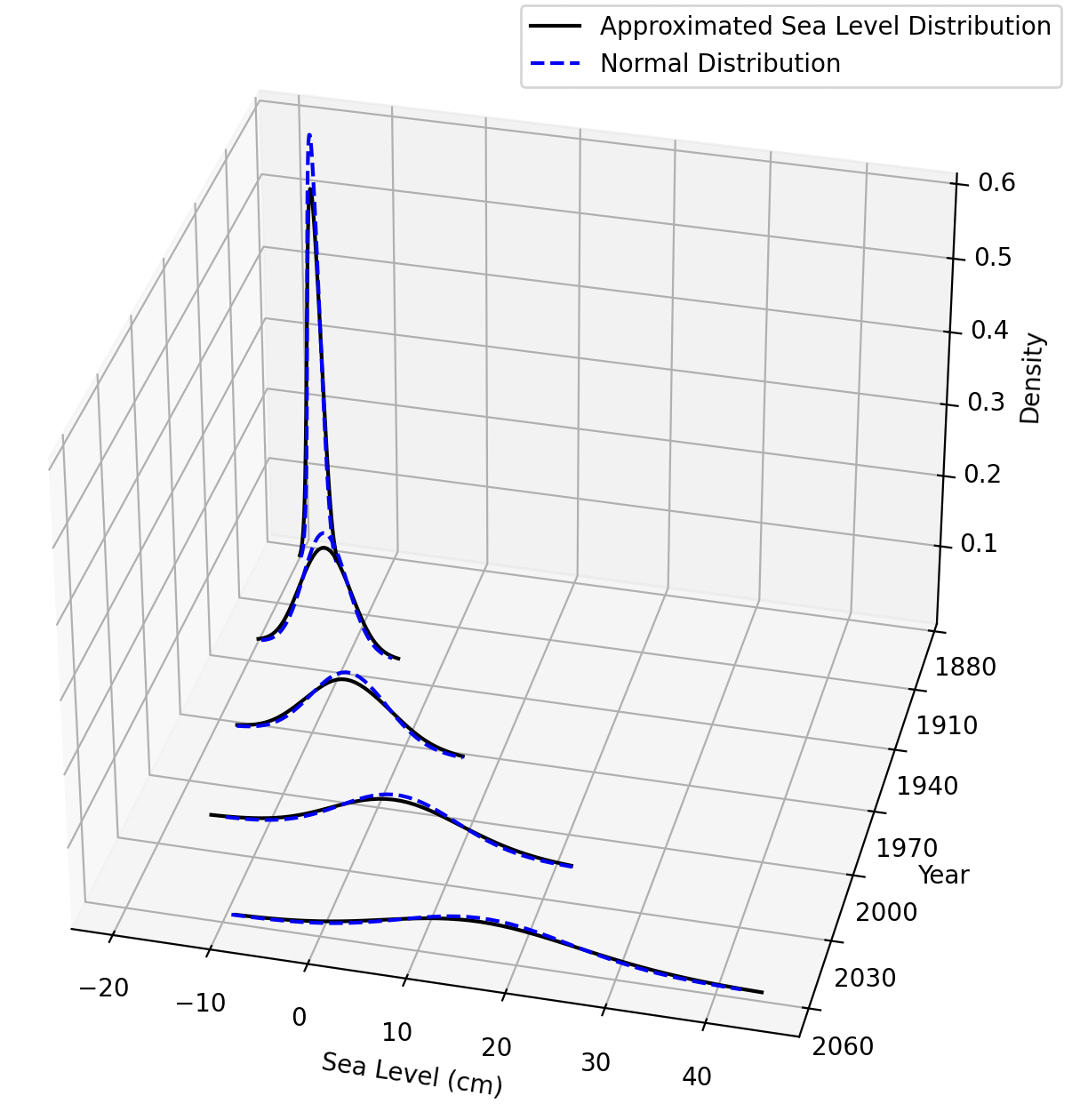}
\caption{Comparison of the approximated sea level distribution against a normal distribution with matching mean and standard deviation. The normal distribution is depicted with a three-standard-deviation range.}

\label{fig:seadis}
\end{figure}

We conducted 10000 simulations comprising 144 time steps each from 1880 to 2053, and the resulting temperature anomaly and sea level distributions are depicted as histograms at every 43 years in Figure \ref{fig:temphist2} and Figure \ref{fig:seahist}.

We calculate the discrete KL-divergence of the latest distribution in the time series in a manner similar to the one-dimensional case. For the temperature anomaly data, the KL-divergence value is 0.0050. For the sea level data, the KL-divergence is 0.0055, suggesting a slightly greater divergence compared to the temperature anomaly data. A reference discrete KL-divergence is calculated between two normal distributions, yielding 0.0021. This indicates that the observed distributions are divergent from the normal distribution. 

From Figure \ref{fig:tempdis2} and Figure \ref{fig:seadis}, it is apparent that the approximated distributions for both temperature anomaly and sea level exhibit flatter characteristics compared to the normal distribution, indicative of lower kurtosis. This flatter spread suggests a less pronounced concentration around the mean, contrasting sharply with the more peaked spread typical of a normal distribution. Additionally, the standard deviation of the distribution of temperature anomaly is 0.45, while that of sea level is 0.63, highlighting a greater variability in sea level fluctuations. 

\section{Statistical Inference}
In this section, we introduce the setup and implementation of the filtering methods. Table \ref{tab:notation} summarizes the notation used in this section.

\begin{table}
  \caption{Notation.}
  \centering
  \setlength{\tabcolsep}{5mm}{
  \begin{tabular}{c p{8cm} c}
  \toprule 
  Parameter & \multicolumn{1}{c}{Description}  &  \\
  \midrule
   $Q$,  $q$  & Process noise variance and standard deviation \\
  $P$,  $p$  & State variance and standard deviation \\
  $R$,  $r$  & Measurement noise variance and standard deviation \\
   $L$ &Dimension of states \\
   $K$ &Dimension of measurement output \\
  $f$ &State governing function \\
  $h$ &Measurement function \\
  $y_{real}$& Data collected from real-world climate system\\
  $y_{obs}$ &The real-world data with measurement noise\\
  $N$& Total number of time steps\\
  $I$& Sample size, equivalent to particle size, ensemble size\\
  $\alpha$ & A scaling parameter that determines the spread of the sigma points around the mean state value ($0 \leq \alpha \leq 1$)\\
  $\beta$ & A non-negative weighting term which can be used to incorporate knowledge of the higher order moments of the distribution ($0 \leq \beta$)\\
  $\kappa$ &  A scaling parameter that is usually set to 0 ($0 \leq \kappa$ to ensure the non-negativity of the variance) &  \\
  $\lambda$ &A composite scaling parameter that equals to $\alpha^2*(L+\kappa)-L$\\
  $K$  & Kalman Gain \\
  $\bm \chi$ &Sigma points and $\chi_i$ is the $i^{th}$ sigma point \\
  $\bm W$ &Weights and $W_i$ is the weight associated with the $i^{th}$ sigma point such $\sum_{i=0}^{2L} W_i = 1$. $W_i$ is calculated according to \eqref{weights} \\
  $\bm \Upsilon$ &Forecast measurement calculated by propagating sigma points though measurement function\\
  
  $\overline{x}$ &  Mean state value\\
  $\overline{y}$ & Mean forecast measurement value\\
  $\hat{x}$ &  Estimated state value\\
  $\hat{y}$ & Estimated forecast measurement value\\
  $\hat{P}$ &  Estimated state variance\\
  $\tilde{x}$ & Resampled particles\\
  $\tilde{w}$ & Normalized weight after resampling\\
  $\tilde{P}$ & State variance after resampling\\

   \bottomrule
  \end{tabular}}
  \label{tab:notation}
\end{table}

Consider the state function $f$ and the measurement function $h$ for nonlinear, non-Gaussian system, the model can be simply expressed as follows:
\begin{align}
x_n &= f(x_{n-1}, n,Q) \\
y_n &= h(x_n, R),
\end{align}
where, in this case, $y_n \in \mathbb{R}^K$ denotes the measurement, $x_n \in \mathbb{R}^L$ the state of the system, $Q$ is the process noise variance, $R$ is the measurement noise variance, and $n$ is the time step. The mappings $f : \mathbb{R}^L \times \mathbb{R} \times \mathbb{R}^L \rightarrow \mathbb{R}^L$ and $h : \mathbb{R}^L \times \mathbb{R}^K\rightarrow \mathbb{R}^K$ represent the state and measurement models. To complete the specification of the model, the prior distribution (at $n = 0$) is denoted by $p(x_0)$. 
The posterior density $p(x_{0:n}|y_{1:n})$, where $x_{0:n} = \{x_0, x_1, \ldots, x_n\}$ and $y_{1:n} = \{y_1, y_2, \ldots, y_n\}$, constitutes the complete solution to the sequential estimation problem. 

Additionally, let $y_{obs}$ denote the observations with measurement noise. $y_{obs,n}= y_{real, n} +\epsilon, \hspace{0.1em} \epsilon \sim \mathcal{N}(0, R)$, where $y_{real}$ is the data we collected from real-world climate system.

\subsection{One-Dimensional Model Setup}

Recall \eqref{discSDE}, we define the state function $f(T_n,n, Q)$ with $n$ from $1880$ to $2024$ (the length of temperature data) as below:
\begin{align}
f(T_n,n,Q) &= T_n +  \frac{1}{C} \left[  \lambda (T_n-T_0) + f \ln\left(\frac{CO_2(n)}{{CO_2}_{\;PI}}\right)\right] \Delta t + \delta, \\
& \text{where} \quad \delta \sim \mathcal{N}(0, Q), \nonumber \\
& \text{and} \quad CO_2(n) = {CO_2}_{\;PI} \times \left(1 + \left(\frac{n - 1850}{220}\right)^3\right). \nonumber
\end{align}
We used $\Delta t =1$ for both one-dimensional and two-dimensional models in our implementation.

Since the output of the state function and the observation is both temperature, the measurement function is an identity function that is given by 
\begin{equation}
 h(T_n, R) = T_n +\epsilon, \hspace{1cm} \text{where} \quad \epsilon \sim \mathcal{N}(0, R) \nonumber \\.
\end{equation}

\subsection{Two-Dimensional Model Setup}

The setup is similar to the one-dimensional model case described above. Recall \eqref{2dmodel}, we define the state function $f((T_n, H_n), n,  Q)$ with $n$ from 1880 to 2021 (the length of sea-level data) and $Q=(Q_1,Q_2)$ as below:

\begin{equation}
\begin{aligned}
f((T_n, H_n), n, Q) &= \left\{
\begin{aligned}
&T_n + (a_{11}T_n + a_{12}H_n + c_1) \Delta t + \delta_1, \\
&H_n + (a_{21}T_n + a_{22}H_n + c_2) \Delta t + \delta_2,
\end{aligned}
\right. \\
& \text{where } \delta_1 \sim \mathcal{N}(0, Q_1) \text{ and } \delta_2 \sim \mathcal{N}(0, Q_2).
\end{aligned}
\label{2dstatefunc}
\end{equation}

The measurement function is dependent on the definition of observed and hidden variables. 
When both variables can be observed, the function is
\begin{equation}
 h((T_n, H_n),R) = (T_n +\epsilon_1, H_n +\epsilon_2) , \hspace{1cm} \text{where} \quad \epsilon_1, \epsilon_2 \sim \mathcal{N}(0, R) \\.
\end{equation}
When only temperature is observed,
\begin{equation}
 h((T_n, H_n),R) = T_n +\epsilon, \hspace{1cm} \text{where} \quad \epsilon \sim \mathcal{N}(0, R)  \\.
\end{equation}
When only sea level is observed,
\begin{equation}
 h((T_n, H_n),R) = H_n+\epsilon, \hspace{1cm} \text{where} \quad \epsilon \sim \mathcal{N}(0, R)  \\.
\end{equation}
The detailed discussion of observed and hidden varibales will be in Section \ref{obshid} .

\subsection{Unscented Kalman Filter}
UKF represents the state distribution by a Gaussian random variable (GRV) \citep{julier1997new}. This state distribution is specified with a minimal set of deterministically chosen sample points. These sample points capture the true mean and variance of the GRV. When propagated through a nonlinear system, they accurately capture the posterior mean and variance up to the 2nd order for any nonlinearity, with errors introduced only in the 3rd and higher orders. To elaborate, we first explain the unscented transformation (UT), a method for calculating the statistics of a random variable undergoing a nonlinear transformation. The UT builds on the principle that approximating a probability distribution is easier than approximating an arbitrary nonlinear function \citep{julier1997new}. Consider propagating a random variable \( x \) (dimension $L$) through a nonlinear function $g$ to yield $y$. Assume \( x \) has mean \( \overline{x} \) and variance \( P_x \). To calculate the statistics of \( y \), we form a matrix \( \bm \chi \) of \( 2L + 1 \) sigma points \( \chi_i \) (with corresponding weights \( W_i \)), according to the following:
\begin{align}
\chi_0 &= \overline{x} \notag\\
\chi_i &= \overline{x} + \left( \sqrt{(L + \lambda)P_x} \right)_i & i &= 1, \ldots , L \notag\\
\chi_i &= \overline{x} - \left( \sqrt{(L + \lambda)P_x} \right)_i & i &= L+1, \ldots , 2L \notag\\
W_0^{(m)} &= \frac{\lambda}{(L + \lambda)} \notag \\
W_0^{(c)} &= \frac{\lambda}{(L + \lambda)} + (1 - \alpha^2 + \beta) \notag \\
W_i^{(m)} = W_i^{(c)} &= \frac{1}{2(L + \lambda)} \quad & i &= 1, \ldots, L \label{weights}
\end{align}
where \( \lambda = \alpha^2(L + k) - L \) is a scaling parameter. \( \alpha \) determines the spread of the sigma points around \( \overline{x} \). \( \kappa \) is a secondary scaling parameter which is usually set to 0, and \( \beta \) is used to incorporate prior knowledge of the distribution of \( x \). \( \left( \sqrt{(L + \lambda)P} \right)_i \) is the \( i \)th row of the matrix square root. These sigma points are propagated through the function,
\begin{align}
\Upsilon_{i} &= g(\chi_{i}) & i &= 0, \ldots , 2L,
\end{align}
and the mean and variance for \( y \) are approximated using a weighted sample mean and variance of the posterior sigma points, which are
\begin{align}
\overline{y} &\approx \sum_{i=0}^{2L} W_i^{(m)} \Upsilon_i \text{  and} \\
P_y &\approx \sum_{i=0}^{2L} W_i^{(c)} [\Upsilon_i - \bar{y}][\Upsilon_i - \bar{y}]^T.
\end{align}
Note that this method differs substantially from general “sampling” methods (e.g., Monte-Carlo methods such as particle filters) which require orders of magnitude more sample points to propagate an accurate and possibly non-Gaussian distribution of the state. The deceptively simple approach using the UT results in approximations that are accurate to the third order for Gaussian inputs for all nonlinearities. For non-Gaussian inputs, approximations are accurate to at least the second-order, with the accuracy of third and higher-order moments determined by the choice of \( \alpha \) and \( \beta \) \citep{wan2000unscented}.

The UKF is a straightforward application of the scaled unscented transformation to the recursive estimation of state variables in a nonlinear system, aiming to achieve minimum mean-square-error (MMSE) in the predictions. The complete UKF algorithm that updates the mean $\overline{x}$ and variance $P$ of the Gaussian approximation to the posterior distribution of the states is given in Algorithm~\ref{UKF}. The initial state variance $P$ is assigned a value of 1 to maintain the effectiveness of Bayesian inference; a value too high would diminish its effectiveness, whereas a value too low may render the use of filtering techniques unnecessary, as deterministic models could suffice.

\begin{algorithm}

\caption{Unscented Kalman Filter}
\begin{algorithmic}[1]

\State \textbf{Inputs:} Measurement function $h$, State function $f$, Process noise variance $Q$, Measurement noise variance $R$, Observations $\{y_{obs,n}\}_{n=1}^{N}$

\State \textbf{Output:} Mean $\overline{x}_n$ and variance $P_n$ at each time step $n$

\State \textbf{Initialize:}
\Indent
\State $\overline{x}_0 = y_0$
\State $P_0 = 1$

\EndIndent
\For{$n = 1$ to $N$},
\State \textbf{1) Calculate sigma points:}
\Indent
\State $\bm \chi_{n-1} = \left[ \overline{x}_{n-1} \hspace{1em} 
 \overline{x}_{n-1} + \sqrt{(L + \lambda)P_{n-1}} \hspace{1em}  \overline{x}_{n-1} - \sqrt{(L + \lambda)P_{n-1}} \right]$
\EndIndent

\State \textbf{2) Time update:}
\Indent
\State $\chi_{j,n|n-1} = f(\chi_{j,n-1}, n, Q) \hspace{1cm} j=0,\ldots,2L $
\State $\overline{x}_{n|n-1} = \sum_{i=0}^{2L} W_i^{(m)}  \chi_{i,n|n-1}$
\State $P_{n|n-1} = \sum_{i=0}^{2L} W_i^{(c)} [\chi_{i,n|n-1} - \overline{x}_n][\chi_{i,n|n-1} - \overline{x}_n]^T $
\State $\Upsilon_{j,n|n-1} = h(\chi_{j,n|n-1},  0)   \hspace{1cm} j=0,\ldots,2L $
\State $\overline{y}_{n|n-1} = \sum_{i=0}^{2L} W_i^{(m)} \Upsilon_{i,n|n-1}$

\EndIndent

\State \textbf{3) Measurement update:}
\Indent
\State $P_{y_n y_n} = \sum_{i=0}^{2L} W_i^{(c)}[\Upsilon_{i,n|n-1} - \overline{y}_n][\Upsilon_{i,n|n-1} - \overline{y}_{n|n-1}]^T $
\State $P_{x_ny_n} = \sum_{i=0}^{2L} W_i^{(c)}[\chi_{i,n|n-1} - \overline{x}_{n|n-1}][\Upsilon_{i,n|n-1} - \overline{y}_{n|n-1}]^T$
\State $K = P_{x_ny_n}P_{y_n y_n}^{-1}$
\State $\overline{x}_n = \overline{x}_{n|n-1} + K(y_{obs,n} - \overline{y}_{n|n-1})$
\State $P_n = P_{n|n-1} - KP_{y_n y_n}K^T$
\EndIndent
 \EndFor
\end{algorithmic}
\label{UKF}
\end{algorithm}

\subsection{Ensemble Kalman Filter}

The central idea of the EnKF is to propagate an ensemble
of \(I\)  state realizations \(\{x_n^{(i)}\}_{i=1}^I\)
instead of the estimate \(\overline{x}_n\) and the variance
\(P_n\) of the KF \citep{evensen1994sequential, roth2017ensemble}. To represent the error statistics in the forecast step, we assume that at time \(n\), we have an ensemble of \(I\) forecasted state estimates with random sample errors. We denote this ensemble as \(\bm x_{n} \in \mathbb{R}^{I}\), where
$\bm x_{n} = [\begin{array}{cccc}
x^{(1)}_{n} & \cdots & x^{(I)}_{n}
\end{array}],$
and the superscript \(i\) refers to the \(i\)-th forecast ensemble member. Then, the ensemble mean \(\overline{x}_{n} \in \mathbb{R}\) is defined by
\begin{equation}
\overline{x}_{n} = \frac{1}{I} \sum_{i=1}^{I} x^{(i)}_{n}.
\end{equation}
Since the true state \(x_n\) is unknown, we approximate the variance of the measurement $P_{y_n y_n}$ and the cross-variance between the state and the measurements $P_{x_n y_n}$ by using the ensemble members. We define the ensemble error matrix \(Ex_{n} \in \mathbb{R}^{I}\) around the ensemble mean by
\begin{equation}
Ex_{n} =  [\begin{array}{ccc}
x^{(1)}_{n} - \overline{x}_{n} & \cdots & x^{(I)}_{n} - \overline{x}_{n} 
\end{array}],
\end{equation}
and the ensemble of output error \(Ey_{n} \in \mathbb{R}^{I}\) by
\begin{equation}
Ey_{n} = [\begin{array}{ccc}
y^{1}_{n} - \overline{y}_{n} & \cdots & y^{(I)}_{n} -\overline{y}_{n}
\end{array}],
\end{equation}
where $\overline{x}_{n}$ and $\overline{y}_{n}$ is calculated through the following time update step:
\begin{align}
 \bm x_{n|n-1} &= f(\bm x_{n-1},n) \notag\\
 \bm y_{n|n-1} &= h(\bm x_{n|n-1},n) \notag\\
   \overline{x}_{n|n-1} &= \frac{1}{I}\sum_{i=1}^{I} x_{n|n-1}^{(i)}\notag\\
 \overline{y}_{n|n-1} &= \frac{1}{I} \sum_{i=1}^{I} y_{n|n-1}^{(i)}
\end{align}

We then approximate  \(P_{x_n y_n} \) and \(P_{x_n y_n} \) by
\begin{align}
P_{x_n y_n} &= \frac{1}{I-1} Ex_{n}Ey_{n}^T\notag\\
P_{y_n y_n}  &= \frac{1}{I-1}  Ey_{n}Ey_{n}^T
\end{align}
Thus, we interpret the forecast ensemble mean as the best forecast estimate of the state, and the spread of the ensemble members around the mean as the error between the best estimate and the actual state.

We use the classical Kalman filter gain expression and the approximations of the error variances to determine the filter gain \( K\) by
\begin{equation}
K = P_{x_ny_n}P_{y_n y_n}^{-1}.
\end{equation}

The implementation of  EnKF is given in Algorithm~\ref{EnKF}.

\begin{algorithm}
\caption{Ensemble Kalman Filter}
\begin{algorithmic}[1]
\State \textbf{Inputs:} Ensemble size $I$, Measurement function $h$, State function $f$, Process noise variance $Q$, Measurement noise variance $R$, Observations $\{y_{obs,n}\}_{n=1}^{N}$
\State \textbf{Output:} Filtered ensemble $\{\overline{x}_{n}^{(i)}\}_{i=1}^{I}$, and ensemble mean $\overline{x}_{n}$ at each time step $n$

\State \textbf{Initialize:}
\Indent
\State $x_{0}^{(i)}=y_0+ \omega  \hspace{1cm} \omega \sim \mathcal{N}(0, P), \hspace{1em}  i=1 \text{ to } I $ and $P=1$
\EndIndent
\For{$n=1$ to $N$}
    \State \textbf{1) Time update:}
    \Indent
    \For{each ensemble numbers $i=1$ to $I$}
        \State  $x_{n|n-1}^{(i)} = f(x_{n-1}^{(i)},n,Q)$ 
          \State  $y_{n|n-1}^{(i)} = h(x_{n|n-1}^{(i)},R)$
    \EndFor

    \EndIndent
    \State \textbf{2) Measurement update:}
    \Indent
    \State $\overline{x}_{n|n-1} = \frac{1}{I}\sum_{i=1}^{I} x_{n|n-1}^{(i)}$
     \State $\overline{y}_{n|n-1} =\frac{1}{I} \sum_{i=1}^{I} y_{n|n-1}^{(i)}$

\State $P_{x_n y_n} = \frac{1}{I-1} [\bm{x}_{n|n-1} - \overline{x}_{n|n-1} \bm{1}][(\bm{y}_{n|n-1} - \overline{y}_{n|n-1} \bm{1}]^T
$
\State $P_{y_n y_n}  = \frac{1}{I-1} [\bm y_{n|n-1} - \overline{y}_n \bm{1}][\bm y_{n|n-1} - \overline{y}_{n|n-1} \bm{1}]^T$

\State $K = P_{x_ny_n}P_{y_n y_n}^{-1}$
\State $\overline{x}_n = \overline{x}_{n|n-1} + K(y_{obs,n} - \overline{y}_{n|n-1})$
\State $\overline{x}_{n} = \frac{1}{I}\sum_{i=1}^{I} x_{n|n-1}^{(i)}$
\EndIndent
\EndFor

\end{algorithmic}
\label{EnKF}
\end{algorithm}

\subsection{Unscented Particle Filter}
The UPF combines the strengths of the UKF and particle filtering to create a robust framework for sequential estimation in complex scenarios \citep{van2000unscented}. Within the UPF, the UKF is utilized to generate proposal distributions, leveraging its adept handling of nonlinearities and its capacity to deal with heavy-tailed noise and outliers. The UKF's strength lies in its precise estimation of the posterior variance, up to the third order, which allows for a better alignment with the true state variance. This makes it highly effective for generating more accurate proposal distributions in the particle filter framework. 

Moreover, unlike traditional KFs that may rely on Gaussian approximations, the UPF's particle filtering component excels at managing non-Gaussian distributions. This integration enables the UPF to maintain a close alignment with the true posterior distribution, significantly enhancing its performance over standalone UKF implementations and other filtering methods. This attribute is crucial in real-world applications where the state distributions are often non-Gaussian and multimodal, making the UPF a powerful tool for tracking and predicting states in dynamic systems.

The implementation of the UPF is in Algorithm~\ref{alg:UPF}. Note in the pseudo-code, $\overline{x}_{n}^{(i)}$ represent the mean value of the propagated sigma points of the $i^{th}$ particle at time $n$. \( \hat{x}_{n} \) represents the estimated state at time \( n \). $\tilde{x}_{n}^{(i)}$ represents the resampled particles with their normalized importance weights $\tilde{w}_{n}^{(i)}$ and corresponding variance $\tilde{P}_{n}^{(i)}$  at time \( n \).

\begin{algorithm}
\caption{Unscented Particle Filter}
\label{alg:UPF}
\begin{algorithmic}[1]
\State \textbf{Inputs:} Measurement function $h$, State function $f$, Process noise variance $Q$, Measurement noise variance $R$, Observations $\{y_{obs,n}\}_{n=1}^{N}$
\State \textbf{Output:} $\mathbb{E}[x_{n}] = \sum_{i=0}^{I} \tilde{w}_{n}^{(i)} * \tilde{x}_{n}^{(i)} \hspace{1em} \text{for } n=1,\ldots,N$

\State \textbf{Initialization:} $t = 0$
\Indent
    \State \textbf{for }$i = 1, \ldots, I$, set $P_0^{(i)} = 1$ and draw particles $x_0^{(i)}$ from the prior $p(x_0) =\mathcal{N}(y_0, P_0^{(i)})$, and set $\overline{x}_0^{(i)} = E[x_0^{(i)}] $

\EndIndent

\For{$n = 1, 2, \ldots, N$}
        \State \textbf{\rom{1}.  Importance sampling step}
        \Indent
        \For{$i = 1, \ldots, I$}
            \State\textbf{a. Update the particles with the UKF:}
            \Indent 
            \State \textbf{1) Calculate sigma points:}
            \Indent
            \State $\bm \chi_{n-1}^{(i)} = \left[ \overline{x}_{n-1}^{(i)} \hspace{1em} \overline{x}_{n-1}^{(i)} + \sqrt{(L + \lambda)P_{n-1}^{(i)}} \hspace{1em} \overline{x}_{n-1}^{(i)} - \sqrt{(L + \lambda)P_{n-1}^{(i)}} \right]$
            \EndIndent

            \State \textbf{2) Time update:}
            \Indent
            \State $ \chi_{j,n|n-1}^{(i)} = f(\chi_{j, n-1}^{(i)}, n, Q) \hspace{1cm} j=0,\ldots,2L $
            \State $\overline{x}_{n|n-1}^{(i)} = \sum_{j=0}^{2L} W_j^{(m)}  \chi_{j,n|n-1}^{(i)}$
            \State $P_{n|n-1}^{(i)} = \sum_{j=0}^{2L} W_j^{(c)} [\chi_{j,n|n-1}^{(i)} - \overline{x}_n^{(i)}][\chi_{j,n|n-1}^{(i)} - \overline{x}_n^{(i)}]^T$
            \State $\Upsilon_{j, n|n-1}^{(i)} = h(\chi_{j, n|n-1}^{(i)}, 0) \hspace{1cm} j=0,\ldots,2L $
            \State $\overline{y}_{n|n-1}^{(i)} = \sum_{j=0}^{2L} W_j^{(m)} \Upsilon_{j,n|n-1}^{(i)}$

            \EndIndent

            \State \textbf{3) Measurement update:}
            \Indent
            \State $P_{y_n y_n} = \sum_{j=0}^{2L} W_j^{(c)}[\Upsilon_{j,n|n-1}^{(i)} - \overline{y}_{n|n-1}^{(i)}][\Upsilon_{j,n|n-1}^{(i)} - \overline{y}_{n|n-1}^{(i)}]^T$
            \State $P_{x_n y_n} = \sum_{j=0}^{2L} W_j^{(c)}[\chi_{j,n|n-1}^{(i)} - \overline{x}_{n|n-1}^{(i)}][\Upsilon_{j,n|n-1}^{(i)} - \overline{y}_{n|n-1}^{(i)}]^T$
            \State $K = P_{x_ny_n}P_{y_n y_n}^{-1}$
            \State $\overline{x}_n^{(i)} = \overline{x}_{n|n-1}^{(i)} + K(y_{obs,n} - \overline{y}_{n|n-1}^{(i)})$
            \State $\hat{P}_n = P_{n|n-1} - KP_{y_n y_n}K^T$
            \EndIndent
         \EndIndent
       
         \State \textbf{b. Sample  } $\hat{x}_n^{(i)} \sim q(x_n^{(i)}|x_{0:n-1}^{(i)}, y_{obs,1:n}) = \mathcal{N}(\overline{x}_n^{(i)}, \hat{P}_n^{(i)})$
        \State \textbf{c. Set  } $\hat{x}_{0:n}^{(i)} \triangleq [x_{0:n-1}^{(i)}, \hat{x}_n^{(i)}]$ and $\hat{P}_{0:n}^{(i)} \triangleq [P_{0:n-1}^{(i)}, \hat{P}_n^{(i)}]$
        \EndFor
        \State \textbf{for} $i = 1, \ldots, I$, evaluate the importance weights up to a normalizing constant.
        \Indent
        \State $w_n^{(i)} \propto \frac{p(y_{obs,n} | \hat{x}_n^{(i)}) p(\hat{x}_n^{(i)} | x_{n-1}^{(i)})}{q(\hat{x}_n^{(i)} | x_{0:n-1}^{(i)}, y_{obs,1:n})} $  
        \EndIndent
        \State \textbf{for} $i = 1, \ldots, I$, normalize the importance weights.
        \EndIndent
    
    \State \textbf{\rom{2}. Selection step}
    \Indent
        \State Multiply/Suppress particles $\left(\hat{x}_{0:n}^{(i)}, \hat{P}_{0:n}^{(i)}\right)$ with high/low importance weights $\tilde{w}_{n}^{(i)}$, \State respectively, to obtain $I$ random particles $\left(\tilde{x}_{0:n}^{(i)}, \tilde{P}_{0:n}^{(i)}\right)$.
    \EndIndent
\EndFor
\end{algorithmic}

\end{algorithm}

\subsubsection{Resampling Methods}
We implemented two types of resampling methods multinomial resampling and systematic resampling \citep{li2015resampling}, and we use primarily systematic resampling for its stability.

\textbf{Multinomial resampling.} 
Multinomial resampling is also referred to as simple random resampling. The core idea of multinomial resampling \citep{gordon1993novel} is to generate independently $I$ random numbers, $u_n^{(i)}$  from the uniform distribution on $(0, 1)$ and use them to select particles from $\bm x_n$. In the $i^{th}$ selection, the particle $x_n^{(m)}$ is chosen when the following condition is satisfied:
\begin{equation}
Q_n^{(m-1)} < u_n^{(i)} \leq Q_n^{(m)},\label{Q}
\end{equation}
where
\begin{equation}
Q_n^{(m)} = \sum_{k=1}^{m} w_n^{(k)}.
\end{equation}

Here, $Q_n^{(m)}$ is the cumulative sum of the weights up to the $m$-th particle at time $n$ and $w_n^{(k)}$ represents the weight of the $k$-th particle at time $n$. Thus, the probability of selecting $x_n^{(m)}$ is the same as that of $u_n^{(i)}$ being in the interval bounded by the cumulative sum of the normalized weights as shown in \eqref{Q}. This sampling scheme satisfies the unbiasedness condition.

\textbf{Systematic resampling.} 
Systematic resampling \citep{kitagawa1996monte}, \citep{carpenter1999improved} exploits the idea of strata, which is dividing the range of probabilities into equal, non-overlapping intervals or "strata." Each particle is then associated with one of these intervals. Now, $u_n^{(1)}$ is drawn from the uniform distribution on $(0, 1/I]$, and the rest of the $u$ numbers are obtained deterministically, i.e.,
\begin{equation}
u_n^{(1)} \sim U(0, \frac{1}{I}],
\end{equation}
\begin{equation}
u_n^{(i)} = u_n^{(1)} + \frac{i - 1}{I}, \quad i = 2, 3, \ldots, I,
\end{equation}
and then the bounding method based on the cumulative sum of normalized weights as shown in  \eqref{Q} is used.

\section{Parameter Optimization and Robustness Testing}
\subsection{Process Noise Variance}\label{sec:prova}

\begin{figure}[ht]
  \centering
  \includegraphics[width=.5\linewidth]{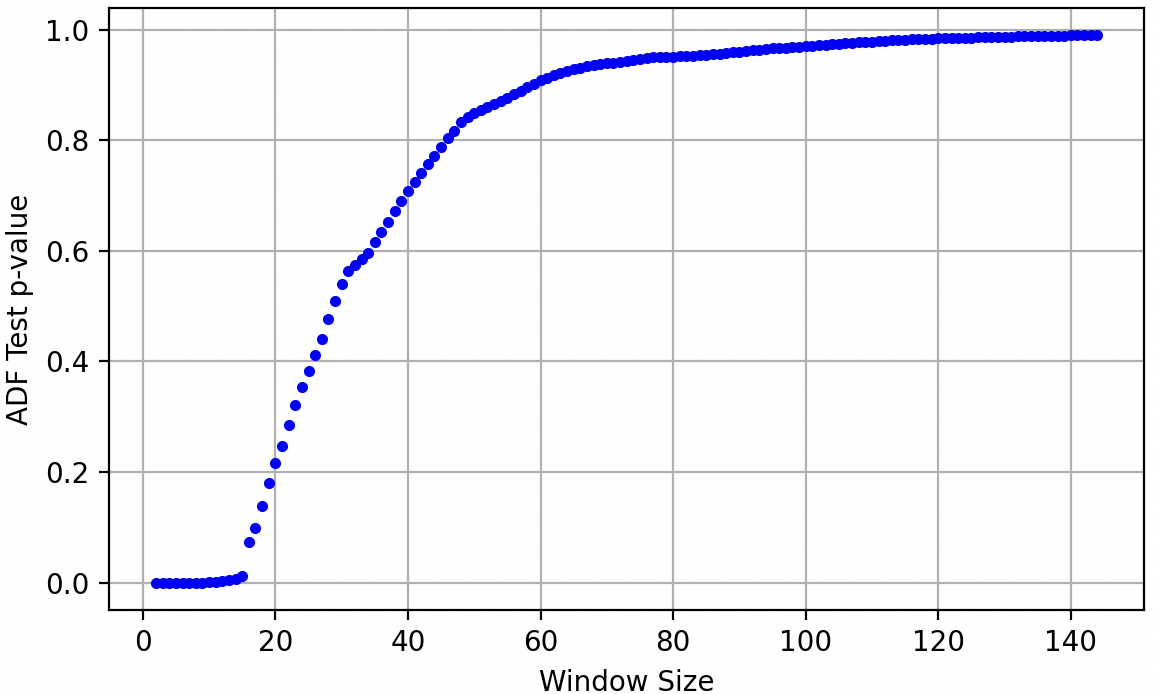}
  \includegraphics[width=.5\linewidth]{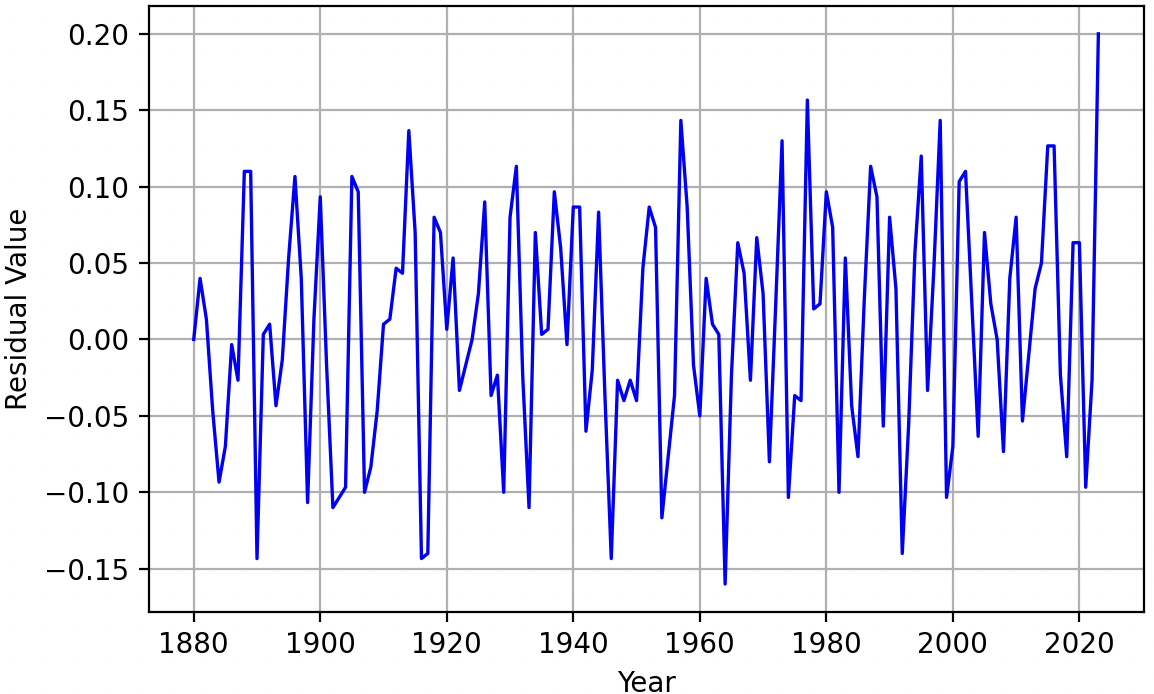}\\

  \caption{Left: Relationship between the Augmented Dickey-Fuller test p-values and various moving average window sizes. Right: Residual values obtained by subtracting the moving average (with the optimal window size of 3) from the original time series data.}

  \label{fig:pval}
\end{figure}

\begin{figure}[ht]
  \centering

  \includegraphics[width=.8\linewidth]{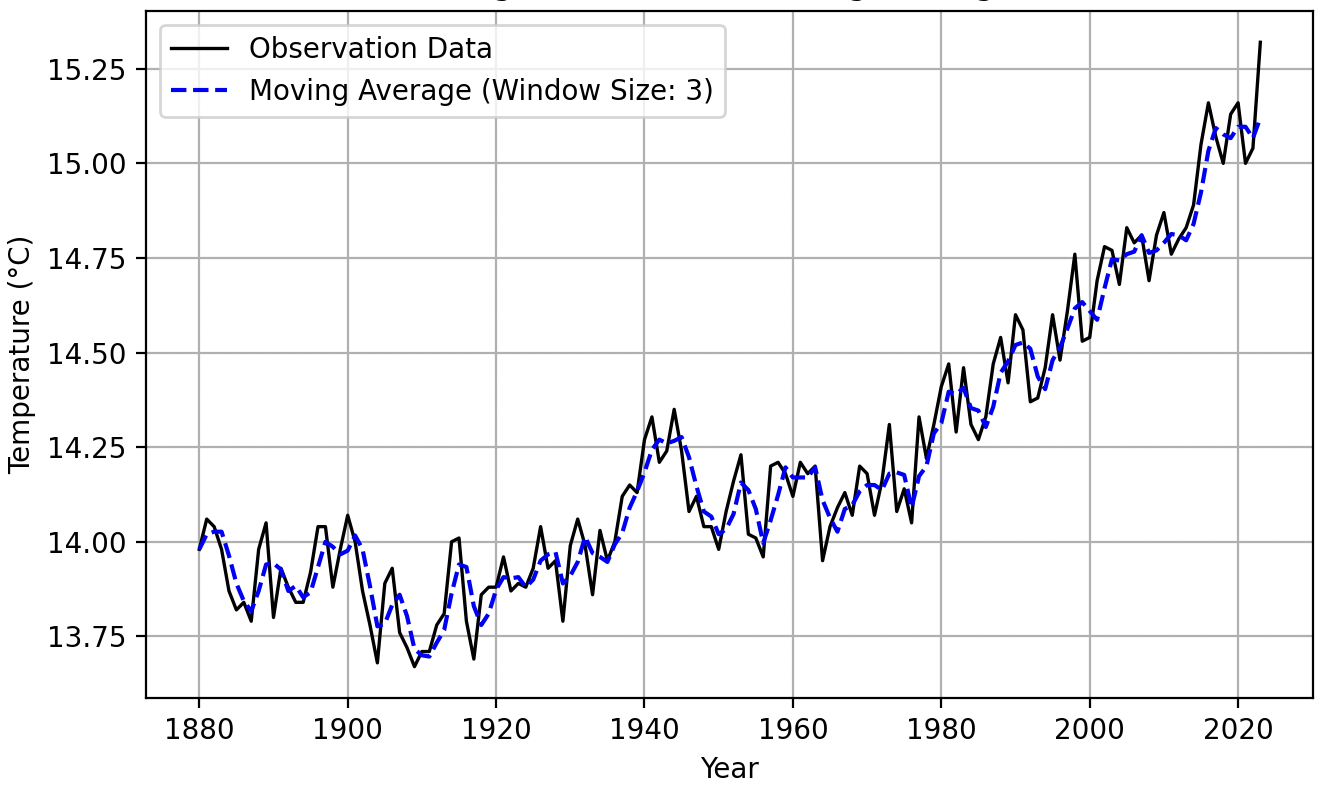}

  \caption{Comparison between the original observation data and the calculated moving average, using an optimal window size of 3, to demonstrate the effect of smoothing on the time series data.}
  \label{fig:moveave}
\end{figure}

In our approach to noise estimation within time series data, we commence by applying a moving average with a window size of \( k \), calculated as:

\begin{equation}
SMA_n = \frac{1}{k} \sum_{i=n-(k-1)}^{t} T_i,
\end{equation}

where \( SMA_n \) represents the simple moving average at time \( n \) and \( T_i \) represents the time series data. The window size \( k \) is chosen through an iterative process aimed at minimizing the p-value from the Augmented Dickey-Fuller (ADF) test, signifying the strongest statistical evidence of stationarity in the residual series. We also check if the minimal p-value is below the significance level (0.05), which infer that the time series is stationary. The resulting p-value corresponding to different window sizes is shown in Figure~\ref{fig:pval} (Left). We find window size 3 yields to the smallest p-value, $1.9\times10^{-23}$, which is also significantly lower than $0.05$.

After determining the best window size, we compute the residual series \( R_n \) by subtracting the moving average from the original time series:

\begin{equation}
R_n = T_n - SMA_n.
\end{equation}

This residual series, shown in Figure~\ref{fig:pval} (Right), signifies the noise component in the data, from which we aim to estimate the noise parameters. To achieve this, we calculate the variance of the residual series \( Var(\bm R) \), assuming the mean of residuals \( \bm R \) to be \( \mu \):

\begin{equation}
Var(\bm R) = \frac{1}{N-1} \sum_{n=1}^{N} (R_n - \mu)^2.
\end{equation}

This variance serves as an estimate for the variance of process noise in our filters $Q \approx Var(\bm R)$. The optimal window size \( k \) thus directly influences our estimate of the noise standard deviation, and consequently, the precision of our particle filter model. It ensures that the estimated noise is not only representative of the true noise affecting the system but is also the optimal estimate under the MMSE criterion.
Figure~\ref{fig:moveave} presents the comparison between the moving average and the original observation data.

For the one-dimensional model described in \eqref{discSDE}, we find the process noise standard deviation $q=0.05$. For the two-dimensional model described in \eqref{2dmodel}, we use $ q= (0.05, 0.3)$.

\subsection{Parameters in UT under High Noise }\label{alpha}

In UT, $\alpha$ controls the size of
the sigma point distribution and should ideally be a small number to avoid sampling non-local effects when the nonlinearities are strong. The parameter \( \alpha \) determines the spread of the sigma points around the mean. A small $\alpha$  in the presence of large measurement noise $R$ can result in a filter that is too confident about its narrow view of the state's probability distribution and that fails to account for the full range of possible states given the noise. This can quickly lead to filter divergence if the true state lies outside of the filter's narrow confidence bounds. Adjusting alpha appropriately or using adaptive methods to adjust the spread of sigma points dynamically can help prevent such issues.

The selection of parameters in UT is attracting increasing attention these days. Many researchers realize that using default values for parameters such as $\alpha$, $\beta$ and $\kappa$ is not optimal. The optimal setting of these parameters is rather complex and currently lacks thorough studies, as it depends on various factors, both explicitly and implicitly. One study that reviews these parameters in detail strongly discourages the use of a small $\alpha$ in the UKF because it results in sigma points being too close to the mean of the state distribution. This proximity limits the model's ability to capture nonlinear effects that occur away from the mean, potentially leading to significant inaccuracies in scenarios where the state distribution is widely spread. The author thus suggests setting $\alpha$ close to 1 \cite{bitzer_ukf}.

Instead using a commonly used small $\alpha$, such as $1 \times 10^{-3}$, we choose $\alpha=0.6$ to ensure the stability under large measurement noise.

\subsection{Particle Variance}

\begin{figure}[ht]
  \centering
  \includegraphics[width=1\linewidth]{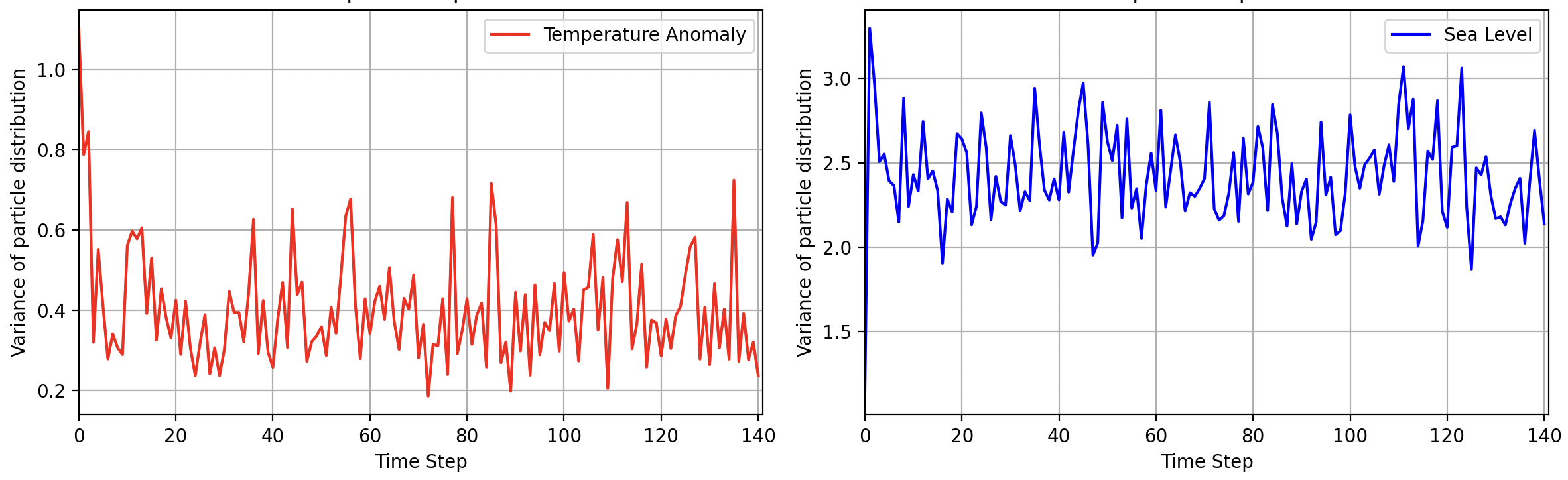}

 \caption{Variance of the posterior particle distribution with measurement noise $r=1$ and sample size 200. Left: Variance of particle distribution over time for temperature anomaly. Right: Variance of particle distribution over time for sea level.}

  \label{fig:variance}
\end{figure}

Figure \ref{fig:variance} presents the variance of the posterior particle distribution for temperature anomaly and sea level with measurement noise $r=1$. The results indicate that the variance remains relatively stable over time, with no significant decrease that would suggest particle degeneracy. The absence of particle degeneracy implies that UPF with systematic resampling is effectively maintaining a diverse set of particles, ensuring an accurate representation of the posterior distribution.

\section{Experiments for One-Dimensional Model}

\begin{table}
  \caption{Setup of the filters for the one-dimensional model.}
  \centering
  \setlength{\tabcolsep}{5mm}{
  \begin{tabular}{lcc}
  \toprule
  Parameter & Value &  Description   \\
  \midrule
  $\alpha$ &0.6 & Scaling parameter \\
  $\beta$ & 2 &Scaling parameter\\
  $\kappa$ & 0 & Scaling parameter  \\

  $q$ & 0.05 & Process noise standard deviation\\
  $N$ & 144 & Total number of time steps \\
$M$ & 100 & Total number of trials \\

  \bottomrule
  \end{tabular}}
  \label{tab:problems1D}
\end{table}

Table \ref{tab:problems1D} presents the setup of filters for the one-dimensional model. We set $\kappa = 0$ to ensure the variance remains non-negative and $\alpha = 0.6$ to effectively manage large noise, as discussed in Section \ref{alpha}. We also use $\beta = 2$. The rationale for choosing $q = 0.05$ is explained in Section \ref{sec:prova}.

In our experiments, we utilize the mean square error (MSE) as the metric for measuring accuracy. The MSE is calculated:
\begin{equation}
MSE = \frac{1}{M}\sum_{j=1}^{M}\frac{1}{N}\sum_{n=1}^{N}(y_{real,n} - x_{n,j})^2,
\label{MSE}
\end{equation}
where $y_{real,n}$ is the real-world observation data and $x_n$ is the filtered estimate. $M$ is the number of trials.

\subsection{Measurement Noise}

We run 100 independent trials using UKF, EnKF and UPF with different measurement noise variances, $R$, with noise $\epsilon$ sampled from normal distributions $\mathcal{N}(0, R)$ for standard deviation $r=0.1, 0.5, 1, 5, \text{ and } 10$.
$MSE$ calculated using \eqref{MSE} is given in Table~\ref{noise}.
\begin{table}[h]
  \caption{Comparison of $MSE$ across UKF, EnKF and UPF under various measurement noise levels, $R$, with noise $\epsilon$ sampled from normal distributions $\mathcal{N}(0, R)$ for $r=0.1, 0.5, 1, 5, \text{ and } 10$. A sample size of 200 was used for both EnKF and UPF. $MSE$ is calculated over 100 trails.}
  \centering
  \setlength{\tabcolsep}{5mm}{
    \begin{tabular}{cccc}
    \toprule 
    Measurement Noise & UKF & EnKF & UPF \\
    \midrule 
    0.1 & 0.028  & 0.012 & 0.0007 \\
    0.5 & 0.129 & 0.018 & 0.006  \\
    1 & 0.433  & 0.038& 0.009 \\
    5 & 0.458 & 0.067  & 0.045 \\
    10 & 0.501 & 0.072 & 0.11  \\
    \bottomrule 
    \end{tabular}
  }\label{noise}
\end{table}

From Table~\ref{noise}, we can observe that UKF performs significantly worse than EnKF and UPF for all noise levels, especially as noise increases. This performance gap can be attributed to the Gaussian assumption underlying both UKF and EnKF. Moreover, UKF employs a deterministic approach using sigma points to approximate the state's mean and variance through nonlinear dynamics, potentially underestimating uncertainty in high noise due to its fixed outcome predictions. Conversely, EnKF adopts a stochastic ensemble method, simulating multiple state instances to incorporate process and measurement noise directly, offering a probabilistic view of future states. This fundamental difference in handling uncertainty makes EnKF generally more adept at managing high noise environments by capturing a wider range of outcomes, compared to the deterministic sigma point method of the UKF.

At a low noise level $r=0.1$, UPF outperforms EnKF substantially, more than 10 times better. For noise levels below 10, UPF outperforms EnKF; however, as noise levels increase to $r=10$, UPF's performance is worse that of EnKF, suggesting UPF's sensitivity to higher noise.
This is because, under high noise conditions, the particle weights in UPF can become highly uneven, leading to a situation where only a few particles dominate the representation of the posterior distribution. This results in a less effective sample from the true posterior, impairing performance. To mitigate this issue, UPF requires a larger particle size to ensure a more diverse set of particles and prevent the concentration of weights on a few particles. Increasing the number of particles helps to better approximate the posterior distribution, maintaining the filter's accuracy in noisy environments, which will be shown next in Section \ref{sec:sample}.

In essence, in low noise level cases, the reliance of UKF and EnKF on Gaussian assumptions introduces an information barrier that constrains their performance under non-Gaussian noise distributions, a limitation that is less pronounced in UPF due to its non-parametric nature. This information barrier arises because both UKF and EnKF are structured around the presumption of Gaussian errors, restricting their ability to accurately represent or adapt to the true complexity of non-Gaussian distribution. This analysis highlights the pivotal role of understanding model assumptions and the characteristics of noise in selecting the most suitable filtering method for precise state estimation.

 \begin{figure}[h!]
  \centering
  \includegraphics[width=.8\linewidth]{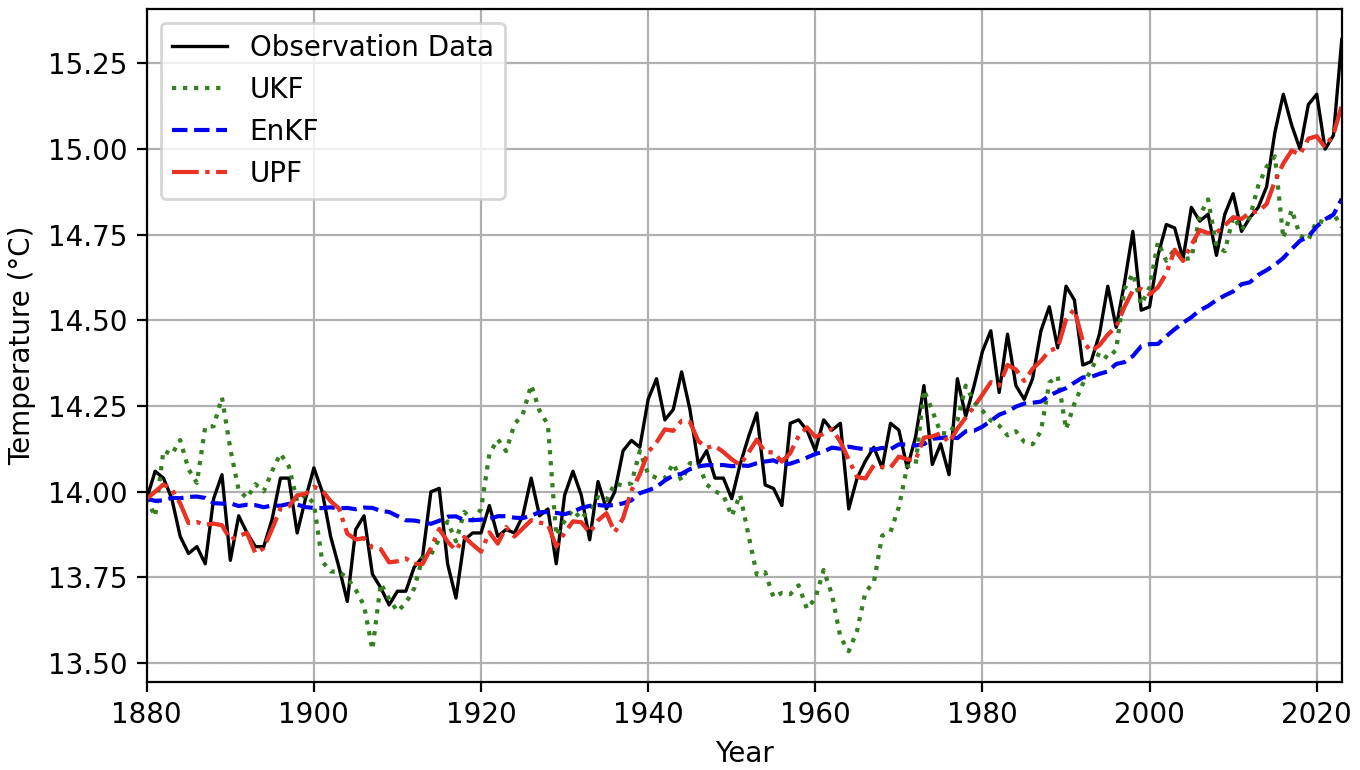}

 \caption{The observation data and the filtered estimates from UKF, EnKF, and UPF under measurement noise $r=1$.}

  \label{fig:1_200}
\end{figure}

\begin{figure}[h!]
  \centering

  \includegraphics[width=.8\linewidth]{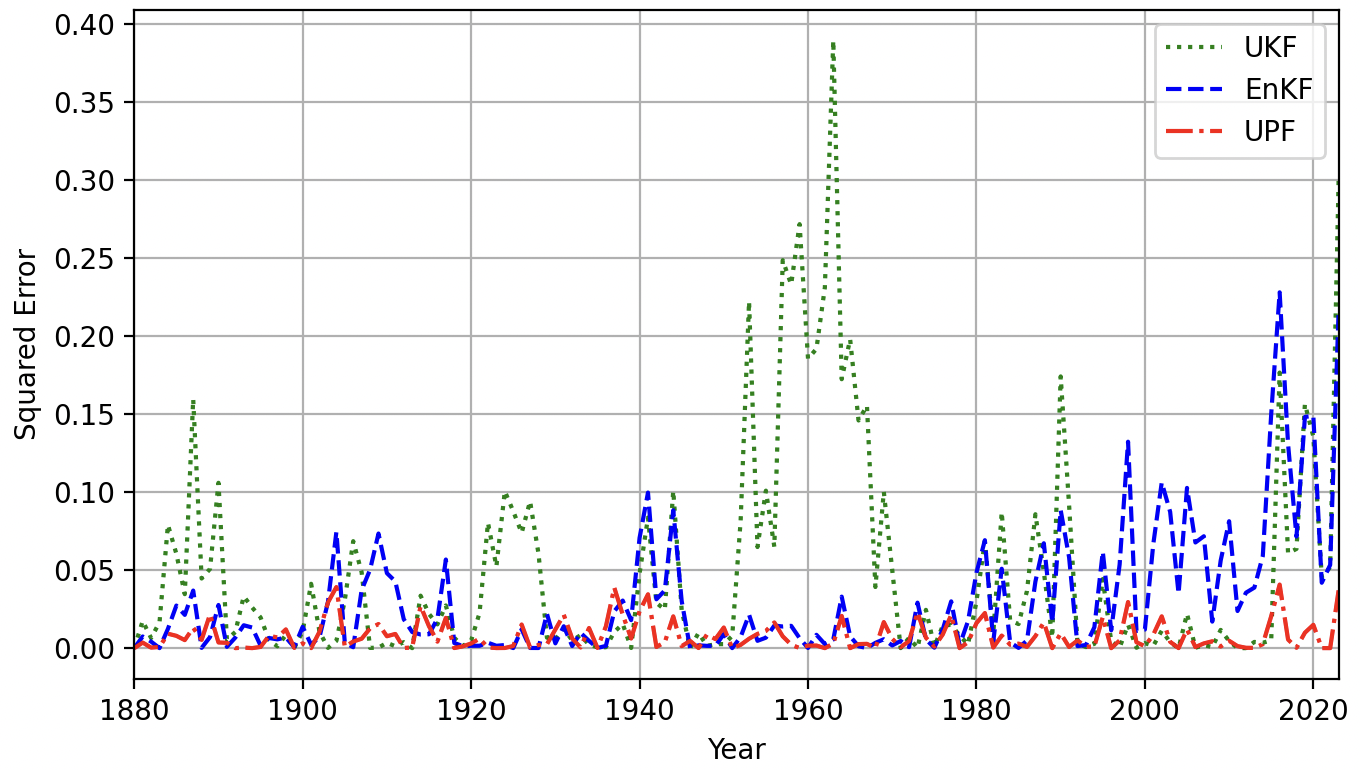}\

 \caption{Squared error of the filtered estimates from UKF, EnKF, and UPF against the observation data under measurement noise $r=1$ calculated using \eqref{se}.}

  \label{fig:1_200error}
\end{figure}

Figure~\ref{fig:1_200} presents observed and filtered temperature data over time with a measurement noise standard deviation of 1. It is evident that UPF most closely matches the observed data. In contrast, UKF displays significant fluctuations, reflecting instability likely due to its deterministic approach. The EnKF output is smoother, suggesting effective noise averaging from its ensemble method. However, EnKF's estimates progressively fall below the observed temperatures as time progresses, which may signal a model bias that underestimates the true temperature increase, indicating that EnKF might rely too heavily on the underlying model.

Figure~\ref{fig:1_200error} displays the squared errors for the three methods, which is calculated as below:

\begin{equation}
    SE_n = \frac{1}{M}\sum_{j=1}^{M}(y_{real,n} - x_{n,j})^2.
    \label{se}
\end{equation}

UPF maintains a consistently low squared error, typically below 0.05 across all years, while EnKF's error escalates notably post 1970. UKF's squared error is markedly higher than the others, with pronounced spikes at several points, indicating its relative instability.

\subsection{Sample Size} \label{sec:sample}
\begin{table}[h]
  \caption{Comparison of $MSE$ across UPF and EnKF with different sample size 10, 50, 100, 200, 500, and 1000 under various measurement noise levels, with $\epsilon$ sampled from normal distributions $\mathcal{N}(0, R)$ for $r=$ 0.5, 5, and 10. $MSE$ is calculated over 100 trials.}
  \centering
  \setlength{\tabcolsep}{3mm}{
    \begin{tabular}{ccccccc}
      \toprule
    Measurement Noise & \multicolumn{2}{c}{0.5} & \multicolumn{2}{c}{5} & \multicolumn{2}{c}{10} \\
    \midrule
    Sample Size & EnKF & UPF & EnKF & UPF & EnKF & UPF \\
    \midrule
    10 & 0.0252 & 0.0297 & 0.088 & 0.819 & 0.087 & 1.600 \\
    50 & 0.0212 & 0.0090 & 0.069 & 0.210 & 0.078 & 0.300 \\
    100 & 0.0199 & 0.0070 & 0.068 & 0.103 & 0.078 & 0.181 \\
    200 & 0.0191 & 0.0056 & 0.067 & 0.045 & 0.070 & 0.110 \\
    500 & 0.0187 & 0.0053 & 0.067 & 0.031 & 0.068 & 0.059 \\
    1000 & 0.0185 & 0.0047 & 0.067 & 0.023 & 0.064 & 0.046 \\
    \bottomrule
    \end{tabular}
  }\label{sample}
\end{table}

 \begin{figure}[ht]
  \centering
  \includegraphics[width=1\linewidth]{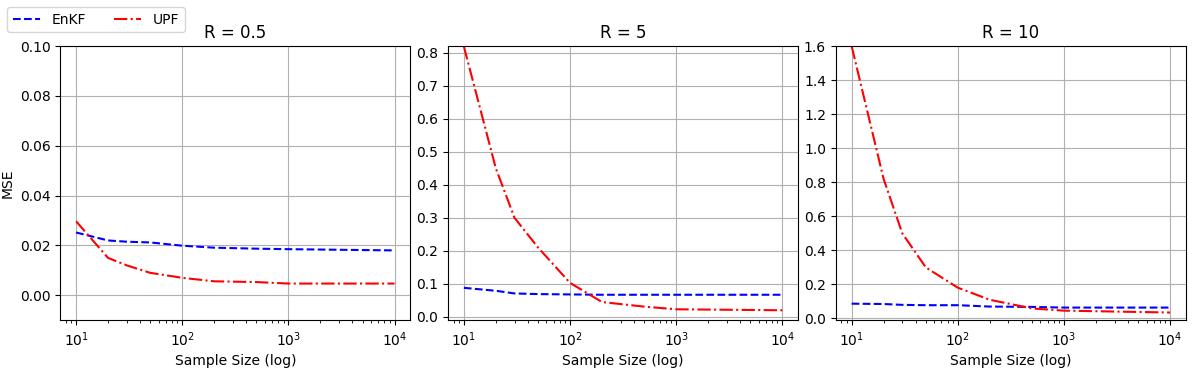}

 \caption{Comparison of $MSE$ across UPF and EnKF with different sample size 10, 20, 30, 50, 100, 200, 500, 1000, and 10000 under various measurement noise levels, with $\epsilon$ sampled from normal distributions $\mathcal{N}(0, R)$ for $r =$ 0.5, 5, and 10. $MSE$ is calculated over 100 trials. Note the x-axis is presented on a logarithmic scale.}

  \label{fig:log}
\end{figure}

Figure~\ref{fig:log} depicts a comparison of the MSE of EnKF and UPF, across a range of sample sizes 10, 20, 30, 50, 100, 200, 500, 1000, and 10000. The MSE is measured under three different measurement noise levels represented by $r =$ 0.5, 5, and 10. 
Table~\ref{sample} records the MSE of EnKF and UPF with sample size 10, 50, 100, 200, 500, and 1000 for $r=$ 0.5, 5, and 10.

For all noise levels, EnKF outperforms UPF when the sample size is 10. However, as the sample size increases, UPF rapidly improves, exhibiting a  logarithmic decline in MSE, ultimately outperforming EnKF for sample sizes of 500 and above. At a noise level of $r=0.5$, both methods demonstrate improvement with increasing sample sizes, yet UPF attains a lower MSE than EnKF for sizes greater than 10. As the noise level increases to $r=5$, the MSE for UPF initially starts much higher than EnKF but decreases more rapidly and eventually falls below the EnKF as the sample size passes 200. At the highest noise level of $r=10$, the UPF begins with a significantly higher MSE compared to EnKF. However, similar to the $r=5$ case, the MSE for UPF reduces sharply as the sample size grows, surpassing the performance of EnKF for sample sizes larger than 500.

These observations suggest that UPF outperforms EnKF as the sample size increases, highlighting UPF's dependence on sample size especially in environments with substantial measurement noise. Conversely, EnKF performs better with smaller sample sizes across all noise levels, with its performance showing marginal improvement as sample sizes increase. This marginal improvement is also due to the information barrier, as EnKF's Gaussian assumptions limit the performance even when the sample size increases. More importantly, this characteristic underscores EnKF's suitability for high-dimensional systems where large sample sizes are impractical due to computational constraints. EnKF is able to deliver stable and relatively accurate results with small ensemble sizes, typically twice the size of the state variables. Even for systems with thousands of states, the literature on EnKF suggests that an ensemble of size 50 to 100 is often adequate \citep{gillijns2006ensemble}. In contrast, as depicted in Figure~\ref{fig:log}, UPF necessitates a significantly larger sample size to attain a similar level of accuracy.

\subsection{Variability }
We run 100 trials and calculated the standard deviation of the errors for all 3 methods.
The the standard deviation of the error is calculated as below.
First, the mean of the error at each timestep \(n\) is calculated as:
\begin{equation}
    \mu_{\text{error}, n} = \frac{1}{M} \sum_{j=1}^{M} (|y_{real,n}-x_{n,j}|) = \frac{1}{M} \sum_{j=1}^{M} e_{n,j},
\end{equation}
where \(x_{n,j}\) is the filtered estimate and \(e_{n,j} = |y_{real,n}-x_{n,j}|\) is the error for the \(j\)th trial at the \(n\)th timestep, and \(M\) is the total number of trials.

Then, the standard deviation of the error at each timestep \(n\), \(\sigma_{\text{error}, n}\), is calculated using the formula:
\begin{equation}
\sigma_{\text{error}, n} = \sqrt{\frac{1}{M} \sum_{j=1}^{M} (e_{n,j} - \mu_{\text{error}, n})^2}.
\end{equation}

\(\sigma_{\text{error}, n}\) thus represents the variability of the error at the \(n\)th timestep across all \(M\) trials.
The overall standard deviation of error is given by

\begin{equation}
\sigma_{\text{error}} = \frac{1}{N} \sum_{i=1}^{N} \sigma_{\text{error}, n}.
\end{equation}

\begin{table}[h]
  \caption{Comparison of $\sigma_{\text{error}}$ across UKF, EnKF and UPF under various measurement noise levels, $R$, with $\epsilon$ sampled from normal distributions $\mathcal{N}(0, R)$ for $r=0.1, 0.5, 1, 5, \text{ and } 10$. A sample size of 200 was used for both EnKF and UPF. $\sigma_{\text{error}}$ is calculated over 100 trials.}
  \centering
  \setlength{\tabcolsep}{5mm}{
    \begin{tabular}{cccc}
    \toprule 
    Measurement Noise & UKF & EnKF & UPF \\
    \midrule 
    0.1 & 0.01  & 0.009 & 0.008 \\
    0.5 & 0.28 & 0.015 & 0.031  \\
    1 & 0.42  & 0.014& 0.058 \\
    5 & 0.66 & 0.027  & 0.170 \\
    10 & 0.63 & 0.028 & 0.250  \\
    \bottomrule 
    \end{tabular}
  }\label{std}
\end{table}

In Table~\ref{std}, We calculate $\sigma_{\text{error}}$ for UKF, EnKF and UPF under various measurement noise levels, $r=$ 0.1, 0.5, 1, 5, and 10, over 100 trials. A sample size of 200 was used for both EnKF and UPF. As seen in Table~\ref{std}, $\sigma_{\text{error}}$ increases for all three methods as measurement noise escalates. Notably, the $\sigma_{\text{error}}$ for UPF surges significantly, whereas the increases for UKF and EnKF are less pronounced after measurement noise reaches $r=5$. This shows that UPF is more sensitive to the increase of measurement, as we discussed in previous section.

 \begin{figure}[h!]
  \centering
  \includegraphics[width=.8\linewidth]{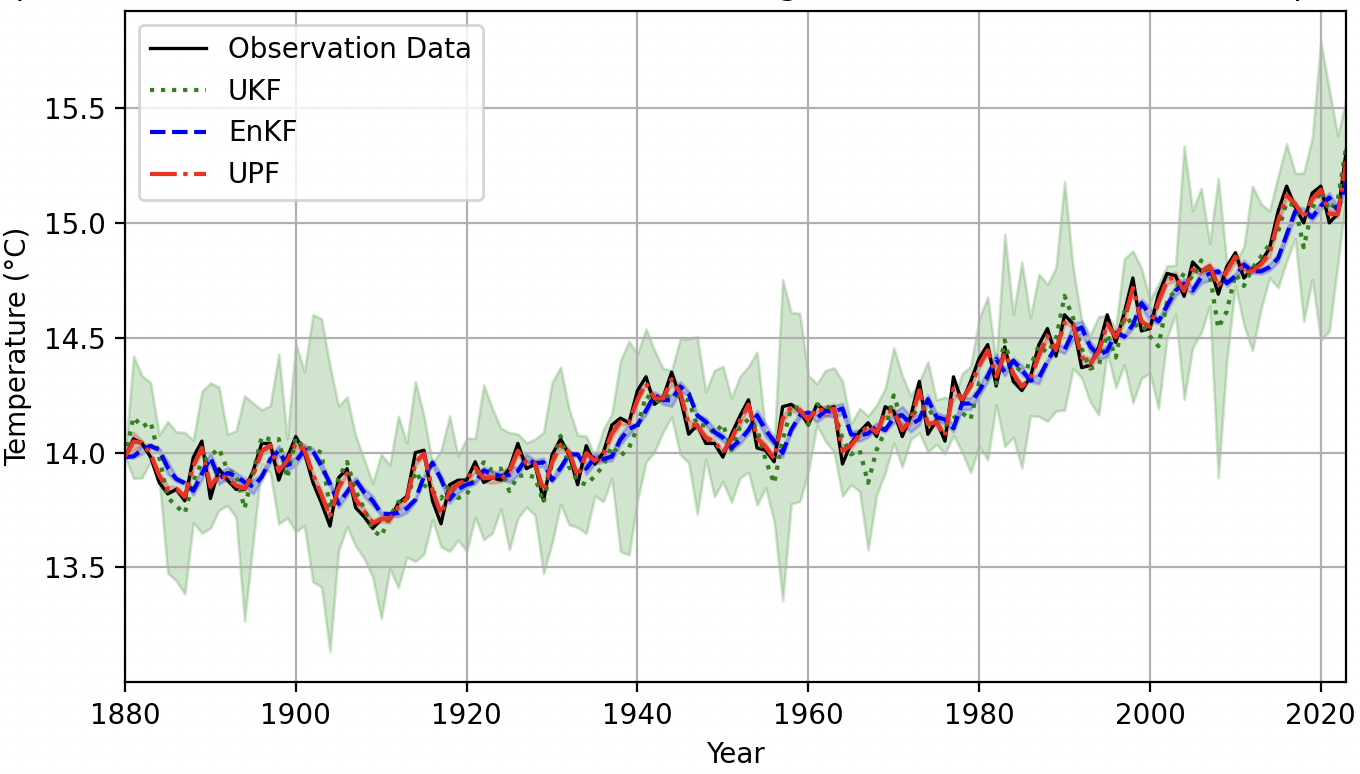}

 \caption{The mean and the confidence intervals (2 standard deviations) of filtered estimates from UKF, EnKF, and UPF under measurement noise $r=0.1$ over 100 trials.}
 \label{fig:CI_0.1}
\end{figure}

 \begin{figure}[h!]
  \centering
  \includegraphics[width=.8\linewidth]{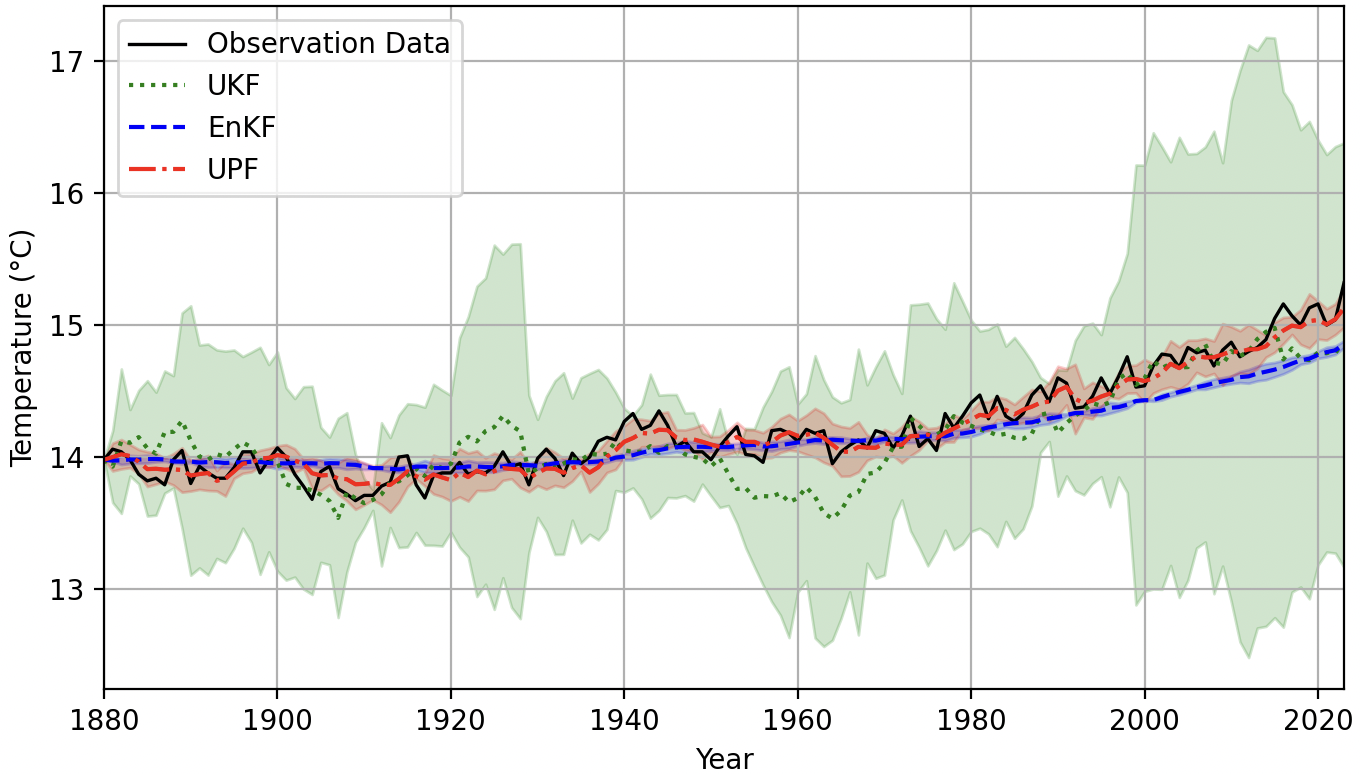}

 \caption{The mean and the confidence intervals (2 standard deviations) of filtered estimates from UKF, EnKF, and UPF under measurement noise $r=1$ over 100 trials.}
  \label{fig:CI_1}
\end{figure}

 \begin{figure}[h!]
  \centering
  \includegraphics[width=.8\linewidth]{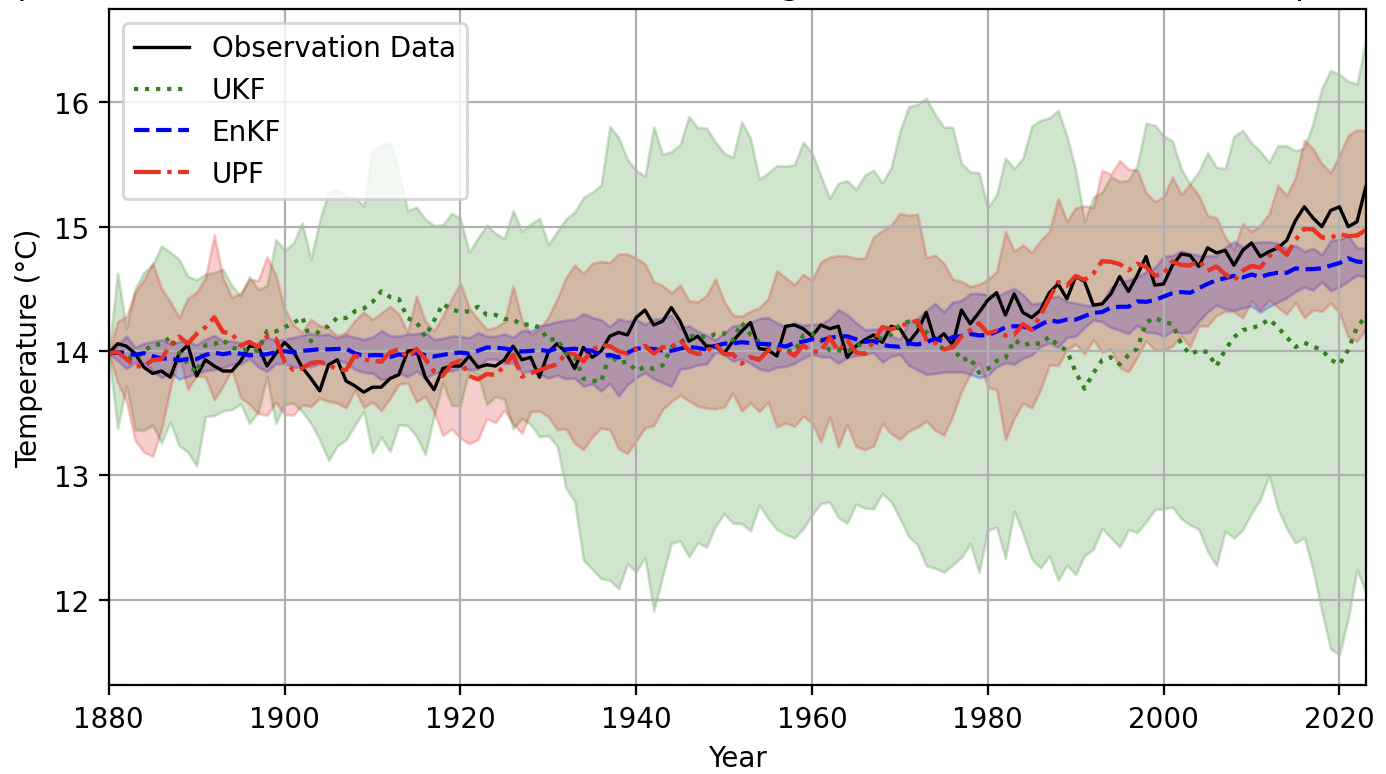}

 \caption{The mean and the confidence intervals (2 standard deviations) of filtered estimates from UKF, EnKF, and UPF under measurement noise $r=10$ over 100 trials. To accommodate the large measurement noise, we increased the process noise standard deviation in the EnKF by a factor of 10, while it remained unchanged in the UKF and UPF. }
  \label{fig:CI_10}
\end{figure}

We also plots the mean of the trials and the confidence intervals (CI) of 2 standard deviations for $r=$ 0.1, 1, and 10 in Figure~\ref{fig:CI_0.1}, \ref{fig:CI_1}, and \ref{fig:CI_10}, respectively. From Table~\ref{std}, Figures~\ref{fig:CI_0.1}, \ref{fig:CI_1}, and \ref{fig:CI_10}, it is evident that the UKF's estimates have the largest $\sigma_{\text{error}}$ and confidence intervals, as indicated by the shaded green area. These intervals suggest a high level of variability in the estimates. The larger $\sigma_{\text{error}}$ implies that the UKF is accommodating a greater range of possible variations in the data. From Figures~\ref{fig:CI_1} and \ref{fig:CI_10}, we also observe that the CI of UKF increases over time, indicating growing variability as time passes. This growing variability could be due to the propagation of sigma points through the nonlinearities in the system. As the sigma points move through these nonlinear transformations at each time step, small errors can accumulate and amplify, leading to a wider prediction spread with each forecast step.

When the measurement noise level is $r=0.1$, the $\sigma_{\text{error}}$ for both the EnKF and UPF are roughly equivalent. As the measurement noise increases, it becomes apparent that the EnKF's estimates have the smallest confidence intervals, represented by the shaded blue area, signifying a lower level of variability in the predictions. This reduced standard deviation suggests that the EnKF might be averaging across multiple ensemble members, which typically decreases the variance of the predicted mean. The confidence intervals for the UPF, shown by the shaded red area, are intermediate in size between those of the UKF and EnKF, indicating that the spread of UPF's predictions is less than UKF's but greater than EnKF's.

To accommodate the large measurement noise when $r=10$, we increased the process noise standard deviation for EnKF by a factor of 10, while it remained unchanged in the UKF and UPF. In Figure~\ref{fig:CI_10}, after 2010, UKF's CI is not covering the observed data, which shows that the prediction range of UKF is too narrow, failing to account for the full range of variability that actually exists in the observed data. UPF's CI encompassing the observation data suggests that the spread of filter's prediction are more representative of the true variability seen in the observed data at that specific time. It indicates that UPF is accounting for the possible states that the system could be in, given the measurement noise.

A CI that captures the observed data points is generally indicative of good filter performance because it means the filter accounts for the true data within its uncertainty bounds. However, it's important to consider both the central tendency (mean prediction) and the variability (CI), since a filter, such as UKF, might have a wider CI that always includes the observed data but whose mean predictions are consistently off-target. For practical applications, such as policy decisions in response to climate change, having a CI that includes the observed data is beneficial as it shows that the filter takes into account the range of possible scenarios that might occur. This is particularly crucial when dealing with non-Gaussian distributions, where it is important to cover the tails of the distribution representing rare events. In the context of climate change, these rare events could be extreme weather conditions, which are vital to consider for accurate and comprehensive risk assessment and planning.

\section{Experiments for Two-dimensional Model}

\begin{table}
  \caption{Setup of the filters for the two-dimensional model.}
  \centering
  \setlength{\tabcolsep}{5mm}{
  \begin{tabular}{lcc}
  \toprule
  Parameter & Value &  Description   \\
  \midrule
  $\alpha$ &0.6 & Scaling parameter \\
  $\beta$ & 2 &Scaling parameter\\
  $\kappa$ & 0 & Scaling parameter  \\

  $q$ & (0.05, 0.3) & Process noise standard deviation\\
  $N$ & 141 & Total number of time steps \\
$M$ & 100 & Total number of trials \\

  \bottomrule
  \end{tabular}}
  \label{tab:problems2D}
\end{table}

Table \ref{tab:problems2D} presents the filter configurations for the tow-dimensional model. We set $\kappa$, $\alpha$, and $\beta$ the same value as in one-dimensional case. The rationale for choosing $ q = (0.05,0.3)$ is explained in Section \ref{sec:prova}. The total number of time steps is 141, corresponding to the length of the sea level data.
In our experiments, we use the normalized mean square error (NMSE) as the accuracy metric to accommodate the varying scales of different variables. The NMSE is calculated:
\begin{equation}
NMSE = \frac{1}{M}\sum_{j=1}^{M}\frac{1}{N}\sum_{n=1}^{N}(\frac{y_{real,n} - x_{n,j}}{max(|\bm y|)})^2,
\label{NMSE}
\end{equation}
where $y_{real,n}$ is the observation data and $x_n$ is the filtered estimate. $M$ is the number of trials. $max(|\bm y|)$ is the maximum absolute value of the data.

\subsection{Sample Size}

\begin{table}[h]
  \caption{Comparison of $NMSE$ of EnKF and UPF estimates with different sample size 10, 50, 100, 200, 500, and 1000 with $\epsilon$ sampled from normal distributions $\mathcal{N}(0, R)$ for $r=$ 0.1. $NMSE$ is calculated over 100 trials. We use both variables as the observed variables.}
    \centering
  \setlength{\tabcolsep}{3mm}{
    \begin{tabular}{ccccc}
      \toprule
    Variable & \multicolumn{2}{c}{Temperature Anomaly} & \multicolumn{2}{c}{Sea Level}  \\
    \midrule
    Sample Size & EnKF & UPF & EnKF & UPF \\
    \midrule
    10 & 0.018 & 0.0094 & $8.50 \times 10^{-4}$ &  $3.64 \times 10^{-4}$  \\
    50 & 0.016 &  0.0063 & $7.88 \times 10^{-4}$ & $2.59\times 10^{-5}$  \\
    100 & 0.015 & 0.0060 &$7.63 \times 10^{-4}$ & $2.30\times 10^{-5}$ \\
    500 & 0.015& 0.0053 & $7.53 \times 10^{-4}$ & $2.11 \times 10^{-5}$ \\
    1000 & 0.015 & 0.0046 &$7.50 \times 10^{-4}$ &  $1.91\times 10^{-5}$ \\
    \bottomrule
    \end{tabular}
  }\label{sample2D}
\end{table}

\begin{table}[h]
  \caption{Comparison of $NMSE$ of temperature anomaly estimates of EnKF and UPF with different sample size 10, 50, 100, 200, 500, and 1000 with $\epsilon$ sampled from normal distributions $\mathcal{N}(0, R)$ for $r=$ 0.1. $NMSE$ is calculated over 100 trials. We use temperature anomaly as the observed variable and the sea level as the hidden variable.}
  \centering
  \setlength{\tabcolsep}{5mm}{
    \begin{tabular}{cccc}
    \toprule 
    Sample Size & EnKF & UPF \\
    \midrule 
    10 & 0.042  & 0.22 \\
    50 & 0.037 & 0.082  \\
    100 & 0.035  & 0.044 \\
    500 & 0.035 & 0.033   \\
    1000 & 0.035 & 0.030   \\
    \bottomrule 
    \end{tabular}
  }\label{sampletemp}
\end{table}

Table \ref{sample2D} and Table \ref{sampletemp} compare the NMSE across two filtering methods, the EnKF and UPF, for different sample sizes when measurement noise $r=0.1$. Notably, when both temperature anomaly and sea level are observed variables, UPF consistently outperforms EnKF across all sample sizes tested. The improvement in performance of UPF is especially noticeable in scenarios with larger sample sizes, indicating better scalability. UPF's non-parametric nature allows it to better capture the complexities and nuances of the data, especially when more data points are available, thus leading to more accurate state estimates.

In scenarios where temperature anomaly is the sole observed variable, both methods performs worse than when both temperature anomaly and sea level are observed. Similar to the one-dimensional case, EnKF achieves a low NMSE with small sample sizes, such as 10 and 50, and the NMSE remains stable after sample sizes exceed 100. This indicates that EnKF is well-suited for high-dimensional systems where large sample sizes are impractical due to computational constraints. The UPF shows better performance than EnKF as the sample size increases to 500, and 1000 samples, demonstrating lower NMSE values in larger datasets. This trend highlights the robustness of UPF in handling non-Gaussian, non-linear systems when provided with more extensive data sets.

\subsection{Observed and Hidden Variables}\label{obshid}

\begin{figure}[!htbp]
  \centering
  \includegraphics[width=.5\linewidth]{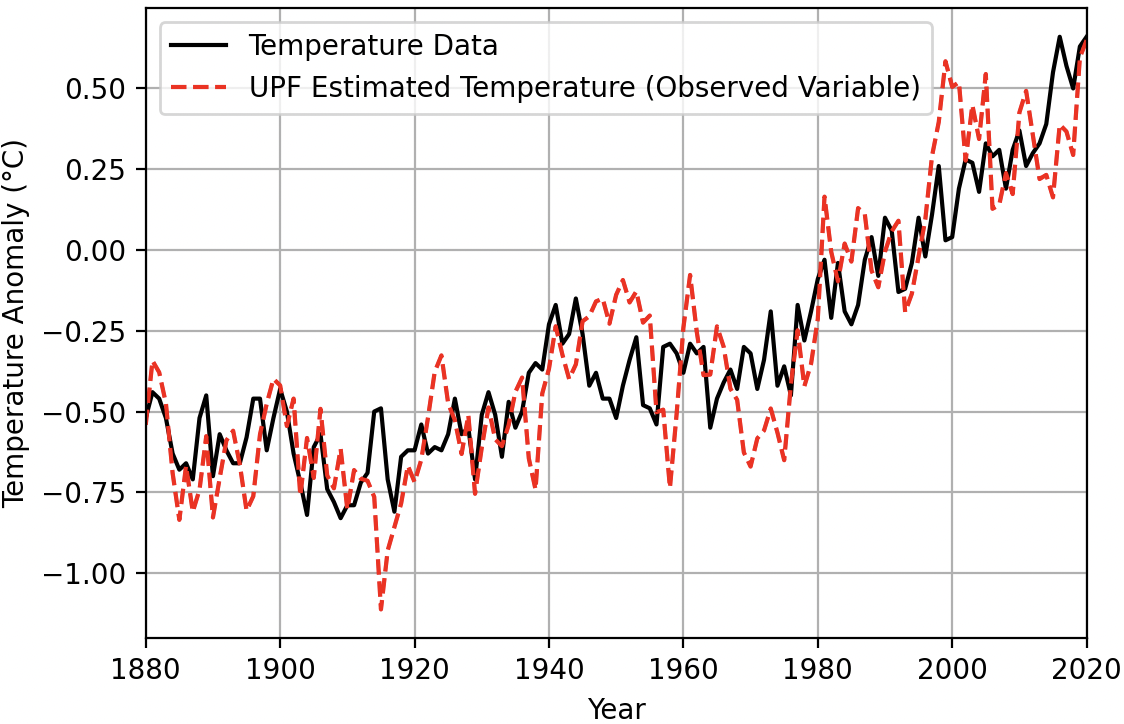}
  \includegraphics[width=.5\linewidth]{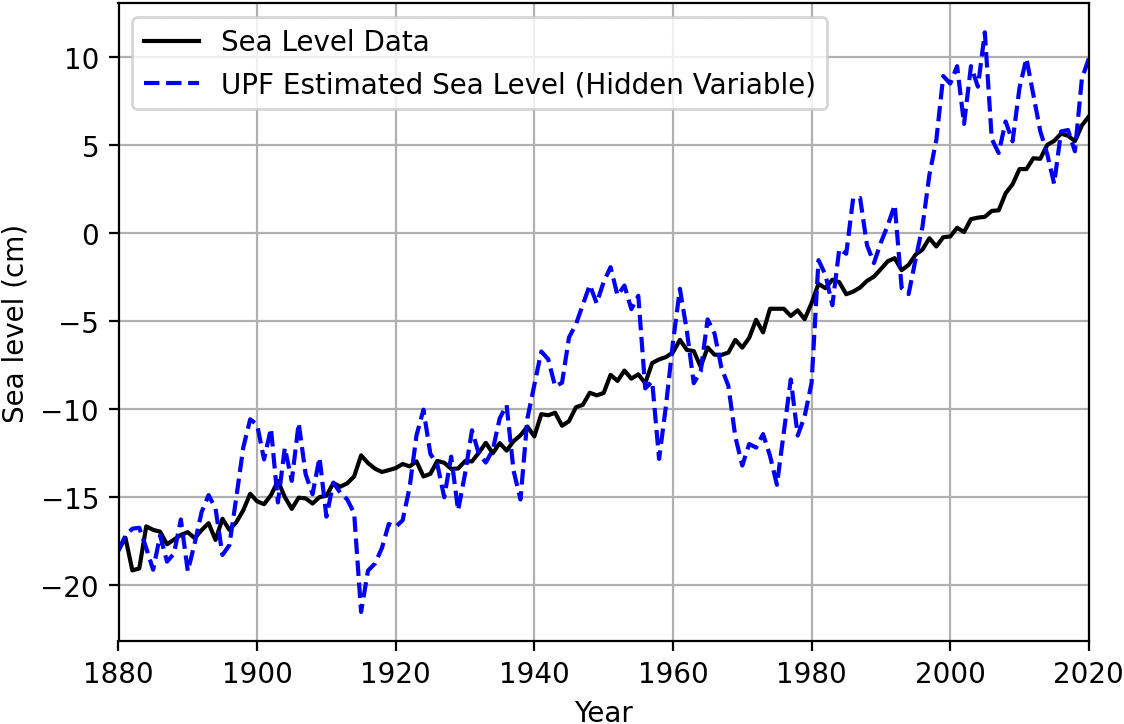}\\

  \caption{
  Left: The temperature anomaly data and the estimated temperature anomaly using UPF. Right: The sea level data and the estimated sea level using UPF. The temperature anomaly is the observed variable and the sea level is the hidden variable. Measurement noise $r = 1$, with a particle size of 200.}

  \label{fig:obstemp}
\end{figure}

\begin{figure}[!htbp]
  \centering
  \includegraphics[width=.5\linewidth]{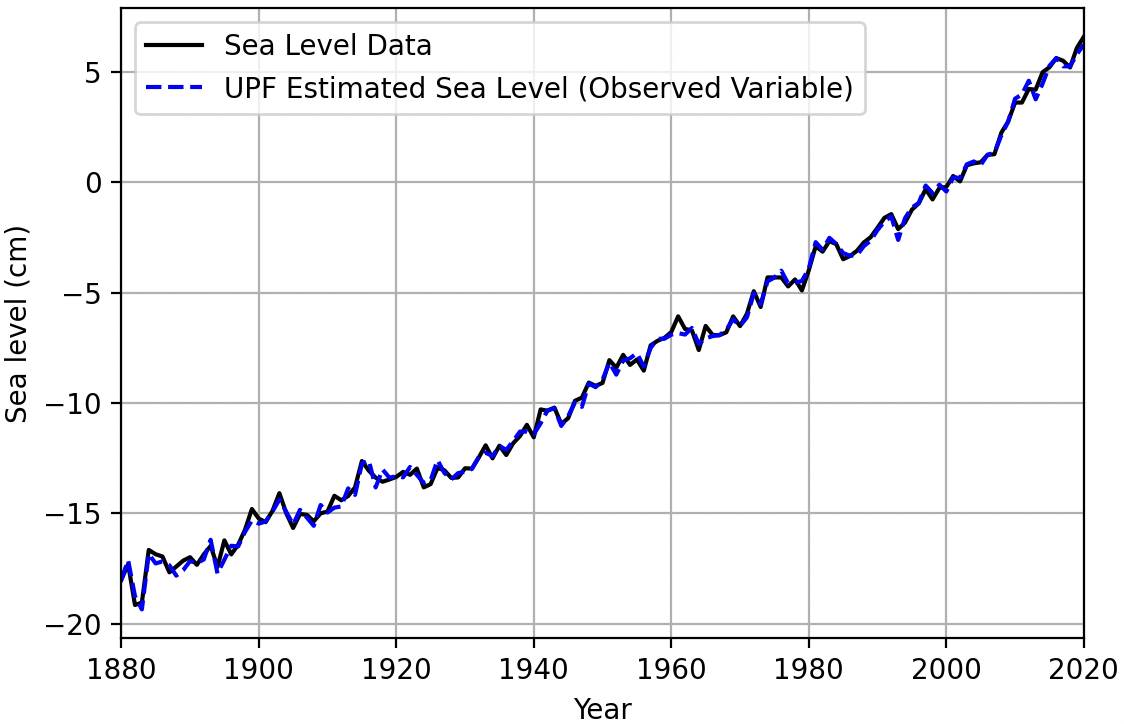}
  \includegraphics[width=.5\linewidth]{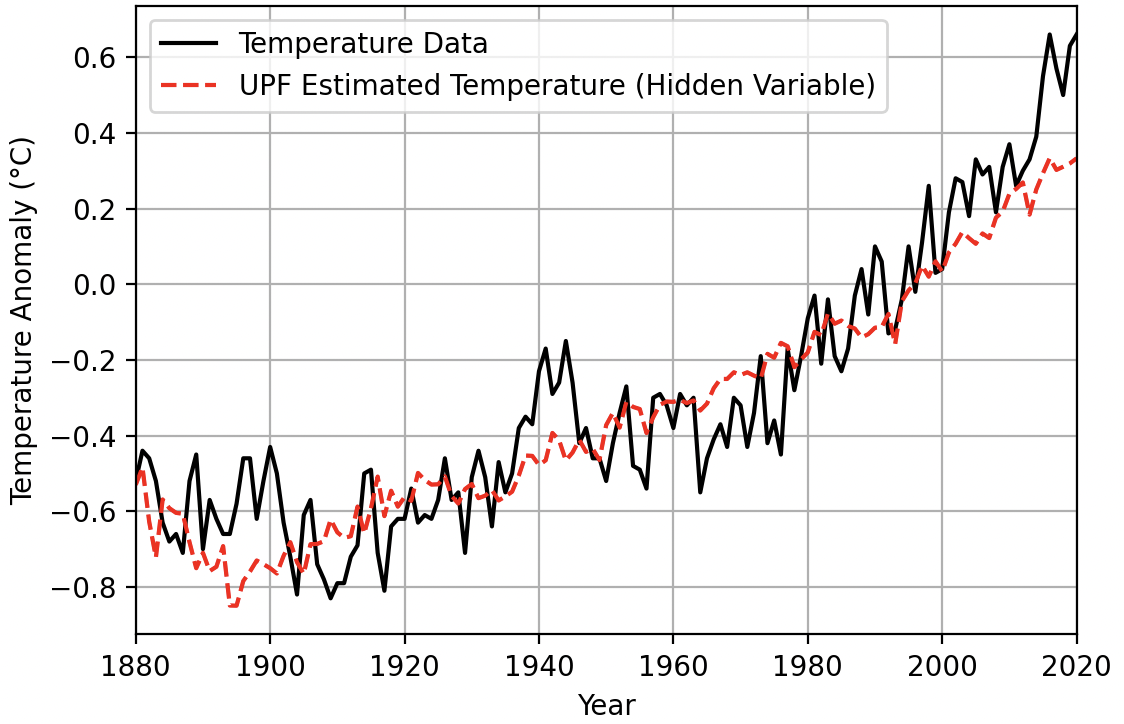}\\

  \caption{Left: The sea level data and the estimated sea level using UPF. Right: The temperature anomaly data and the estimated temperature anomaly using UPF. The sea level is the observed variable and the temperature anomaly is the hidden variable. Measurement noise $r = 1$, with a particle size of 200.}

  \label{fig:obssea}
\end{figure}

\begin{figure}[h]
  \centering
  \includegraphics[width=1\linewidth]{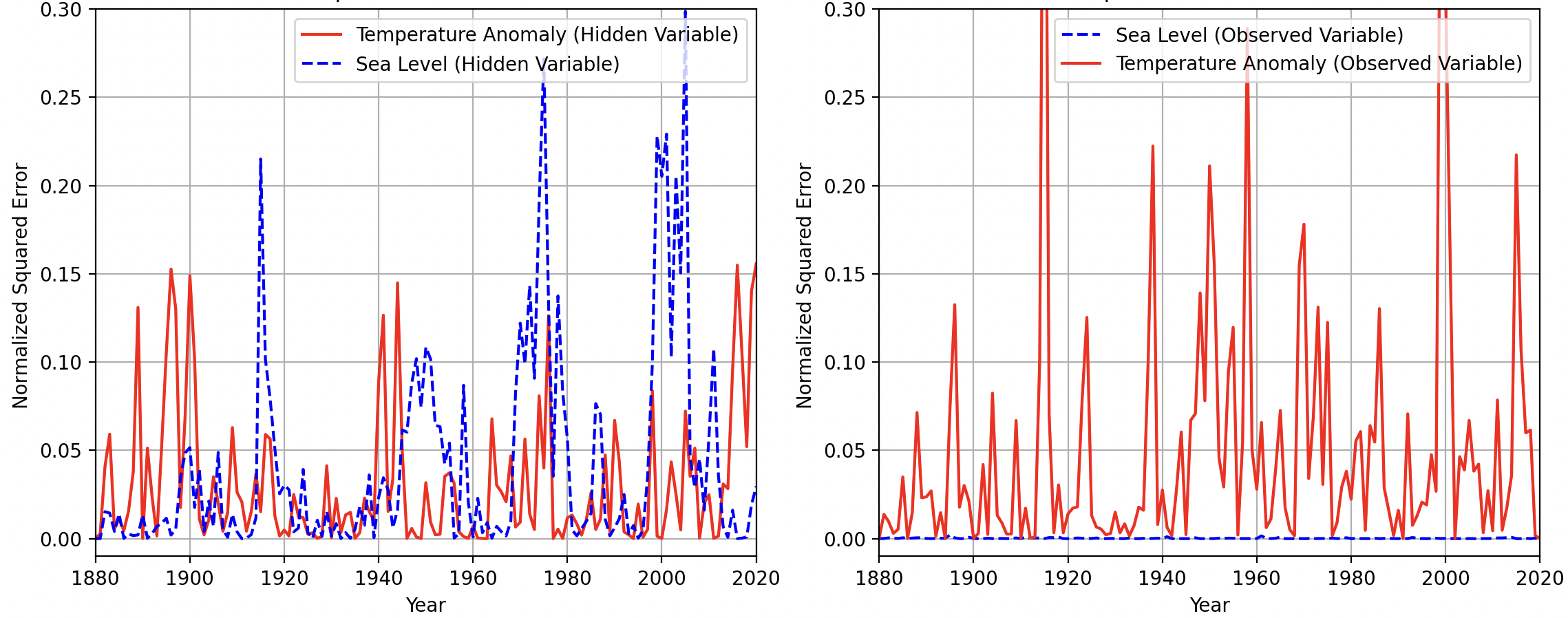}

  \caption{NSE of the filtered estimates from UPF against the observation data under measurement noise $r=1$ with a particle size of 200 calculated using \eqref{NSE}. Left: The NSE of temperature anomaly, when temperature anomaly is the hidden variable and sea level is the observed variable, and the NSE of sea level when sea level is the hidden variable and temperature anomaly is the observed variable. Right: The NSE of sea level when sea level is the observed variable and temperature anomaly is the hidden variable, and the NSE of temperature anomaly, when temperature anomaly is the observed variable and sea level is the hidden variable. }

  \label{fig:obshiderror}
\end{figure}

We test the performance of the UPF with different observed variables under a measurement noise of $r=1$ and a particle size of 200 across 100 trials. Figure \ref{fig:obstemp} displays the results when temperature anomaly is the observed variable and sea level is the hidden variable. Conversely, Figure \ref{fig:obssea} shows the outcomes when sea level is the observed variable and temperature anomaly is the hidden variable.

Figure \ref{fig:obshiderror} plots the normalized squared error (NSE) of the filtered estimates from UPF against the observation data, calculated as below:
\begin{equation}
NSE_n = \frac{1}{M}\sum_{j=1}^{M}(\frac{y_{real,n} - x_{n,j}}{max(|\bm y|)})^2.
\label{NSE}
\end{equation}

The left panel of Figure \ref{fig:obshiderror} presents the NSE for temperature anomaly as the hidden variable and sea level as the observed variable. The NMSE for temperature anomaly is 0.031, and for sea level, it is 0.042. Notably, the NSE values for temperature anomaly consistently remain below 0.16, whereas several peaks in sea level NSE exceed 0.2.

The right panel of Figure \ref{fig:obshiderror} illustrates the NSE for sea level as the observed variable and temperature anomaly as the observed variable. The NMSE for temperature anomaly is 0.051, and for sea level, it is 0.00018. The NSE for sea level is extremely low at all times, not exceeding 0.01, while the NSE for temperature anomaly reaches values over 0.3.

These results indicate that the performance of the filter is superior when sea level is the observed variable, evidenced by lower NMSE and maximum NSE values. The better performance is likely due to the relatively small scale of the measurement error compared to the scale of sea level data, as opposed to the temperature anomaly data. It suggests that the performance of the filter heavily depends on the accuracy with which we can quantify measurement errors. In our experiments, the same measurement error was used for both variables; however, in real-world applications, the scale of error can vary significantly between different types of measurement data.

\section{Conclusion}

\subsection{Summary}
In this study, we proposed a Bayesian framework leveraging the UKF, EnKF, and UPF to address the complexities of non-Gaussian systems in climate modeling. Our evaluation, conducted under various conditions including different measurement noise levels, sample sizes, and observed and hidden variables, provided a thorough analysis of the advantages and limitations of each method. Our findings indicate that in low noise levels and large sample size scenarios, the reliance of UKF and EnKF on Gaussian assumptions introduces an information barrier that constrains their performance under non-Gaussian noise distributions. EnKF is particularly suitable for high-dimensional systems where large sample sizes are impractical due to computational constraints. It delivers stable and relatively accurate results with small ensemble sizes, typically twice the size of the state variables. In contrast, UPF requires a significantly larger sample size to achieve similar accuracy, yet it demonstrates better performance with increasing data. Additionally, UKF's confidence interval often fails to cover the observed data, indicating a prediction range too narrow to account for the true variability. Conversely, UPF's confidence interval encompasses the observed data, showing that its predictions are more representative of the actual variability, even with significant measurement noise. Experiments on different observed and hidden variables suggest that the performance of the filter heavily depends on the accuracy with which measurement errors are quantified. In summary, selecting the appropriate methods and models is crucial for accurate and reliable predictions in data assimilation. Through our study, we aim to provide comprehensive insights for applying these data assimilation techniques in real-world scenarios, highlighting their strengths and limitations to guide practitioners in making informed decisions.

\subsection{Limitation and Future Work}
While the models demonstrated promising results, their performance heavily relies on the accuracy with which we can quantify measurement errors. As evidenced in Section \ref{obshid}, the scale of error varies significantly between temperature and sea level measurements, directly influencing the effectiveness of the models with different observed variables. In future work, we plan to incorporate diverse data sources to enhance the estimation of measurement errors. Additionally, the practical challenges of high-dimensional systems in climate modeling cannot be overlooked. Many models in climate science operate within extremely high-dimensional spaces, where the particle filter framework, despite its advantages, faces substantial computational hurdles. As shown in Section \ref{sample}, although UPF outperformed other methods, the computational cost required to maintain this performance is prohibitive. 

Moving forward, we will explore techniques to reduce computational demands while retaining the robustness of the models. This includes dimensionality reduction strategies such as Principal Component Analysis (PCA) and more efficient algorithmic implementations. Additionally, we plan to investigate Topological Data Analysis (TDA) to capture the underlying topological features of the climate data, such as identifying persistent patterns and structures that might not be evident through traditional statistical methods, as well as new non-parametric statistics, including Quantile Regression Forests (QRF) and non-parametric panel data models.

Moreover, we aim to leverage parametrization and machine learning techniques to enhance scalability and effectiveness in complex systems. This will involve using Recurrent Neural Networks (RNN) and Long Short-Term Memory Networks (LSTM), either independently or in combination with parametrization methods. These efforts are critical for scaling our approaches to more complex systems and ensuring their practical applicability in real-world climate science scenarios.

\bibliographystyle{unsrtnat}
\bibliography{refs}

\begin{thebibliography}{76}
\providecommand{\natexlab}[1]{#1}
\providecommand{\url}[1]{\texttt{#1}}
\expandafter\ifx\csname urlstyle\endcsname\relax
  \providecommand{\doi}[1]{doi: #1}\else
  \providecommand{\doi}{doi: \begingroup \urlstyle{rm}\Url}\fi

\bibitem[{National Oceanic and Atmospheric Administration (NOAA)}(2024)]{climate-gov}
{National Oceanic and Atmospheric Administration (NOAA)}.
\newblock Understanding climate change: Global sea level, 2024.
\newblock URL \url{https://www.climate.gov/news-features/understanding-climate/climate-change-global-sea-level}.
\newblock Accessed on April 18, 2024.

\bibitem[{NASA}(No Date)]{nasa-climate}
{NASA}.
\newblock Vital signs: Global temperature, No Date.
\newblock URL \url{https://climate.nasa.gov/vital-signs/global-temperature/?intent=121}.
\newblock Accessed on April 18, 2024.

\bibitem[Pinsky and Karlin(2010)]{pinsky2010introduction}
Mark Pinsky and Samuel Karlin.
\newblock \emph{An introduction to stochastic modeling}.
\newblock Academic press, 2010.

\bibitem[Palmer et~al.(2005)Palmer, Shutts, Hagedorn, Doblas-Reyes, Jung, and Leutbecher]{palmer2005representing}
TN~Palmer, GJ~Shutts, R~Hagedorn, FJ~Doblas-Reyes, Thomas Jung, and M~Leutbecher.
\newblock Representing model uncertainty in weather and climate prediction.
\newblock \emph{Annu. Rev. Earth Planet. Sci.}, 33:\penalty0 163--193, 2005.

\bibitem[Palmer et~al.(2009)Palmer, Buizza, Doblas-Reyes, Jung, Leutbecher, Shutts, Steinheimer, and Weisheimer]{palmer2009stochastic}
Tim~N Palmer, Roberto Buizza, F~Doblas-Reyes, Thomas Jung, Martin Leutbecher, Glenn~J Shutts, Martin Steinheimer, and Antje Weisheimer.
\newblock Stochastic parametrization and model uncertainty.
\newblock 2009.

\bibitem[Yiou and D{\'e}andr{\'e}is(2019)]{yiou2019stochastic}
Pascal Yiou and C{\'e}line D{\'e}andr{\'e}is.
\newblock Stochastic ensemble climate forecast with an analogue model.
\newblock \emph{Geoscientific Model Development}, 12\penalty0 (2):\penalty0 723--734, 2019.

\bibitem[Franzke et~al.(2015)Franzke, O'Kane, Berner, Williams, and Lucarini]{franzke2015stochastic}
Christian~LE Franzke, Terence~J O'Kane, Judith Berner, Paul~D Williams, and Valerio Lucarini.
\newblock Stochastic climate theory and modeling.
\newblock \emph{Wiley Interdisciplinary Reviews: Climate Change}, 6\penalty0 (1):\penalty0 63--78, 2015.

\bibitem[Palmer(2019)]{palmer2019stochastic}
TN~Palmer.
\newblock Stochastic weather and climate models.
\newblock \emph{Nature Reviews Physics}, 1\penalty0 (7):\penalty0 463--471, 2019.

\bibitem[Ewald et~al.(2004)Ewald, Penland, and Temam]{ewald2004accurate}
Brian Ewald, Cecile Penland, and Roger Temam.
\newblock Accurate integration of stochastic climate models with application to el ni{\~n}o.
\newblock \emph{Monthly weather review}, 132\penalty0 (1):\penalty0 154--164, 2004.

\bibitem[Wilks and Wilby(1999)]{wilks1999weather}
Daniel~S Wilks and Robert~L Wilby.
\newblock The weather generation game: a review of stochastic weather models.
\newblock \emph{Progress in physical geography}, 23\penalty0 (3):\penalty0 329--357, 1999.

\bibitem[Hasselmann(1976)]{hasselmann1976stochastic}
Klaus Hasselmann.
\newblock Stochastic climate models part i. theory.
\newblock \emph{tellus}, 28\penalty0 (6):\penalty0 473--485, 1976.

\bibitem[Majda et~al.(1999)Majda, Timofeyev, and Vanden~Eijnden]{majda1999models}
Andrew~J Majda, Ilya Timofeyev, and Eric Vanden~Eijnden.
\newblock Models for stochastic climate prediction.
\newblock \emph{Proceedings of the National Academy of Sciences}, 96\penalty0 (26):\penalty0 14687--14691, 1999.

\bibitem[Majda et~al.(2001)Majda, Timofeyev, and Vanden~Eijnden]{majda2001mathematical}
Andrew~J Majda, Ilya Timofeyev, and Eric Vanden~Eijnden.
\newblock A mathematical framework for stochastic climate models.
\newblock \emph{Communications on Pure and Applied Mathematics: A Journal Issued by the Courant Institute of Mathematical Sciences}, 54\penalty0 (8):\penalty0 891--974, 2001.

\bibitem[Tingley et~al.(2012)Tingley, Craigmile, Haran, Li, Mannshardt, and Rajaratnam]{tingley2012piecing}
Martin~P Tingley, Peter~F Craigmile, Murali Haran, Bo~Li, Elizabeth Mannshardt, and Bala Rajaratnam.
\newblock Piecing together the past: statistical insights into paleoclimatic reconstructions.
\newblock \emph{Quaternary Science Reviews}, 35:\penalty0 1--22, 2012.

\bibitem[Budyko(1969)]{budyko1969effect}
Mikhail~I Budyko.
\newblock The effect of solar radiation variations on the climate of the earth.
\newblock \emph{tellus}, 21\penalty0 (5):\penalty0 611--619, 1969.

\bibitem[Sellers(1969)]{sellers1969global}
William~D Sellers.
\newblock A global climatic model based on the energy balance of the earth-atmosphere system.
\newblock \emph{Journal of Applied Meteorology and Climatology}, 8\penalty0 (3):\penalty0 392--400, 1969.

\bibitem[North(1975{\natexlab{a}})]{north1975theory}
Gerald~R North.
\newblock Theory of energy-balance climate models.
\newblock \emph{Journal of Atmospheric Sciences}, 32\penalty0 (11):\penalty0 2033--2043, 1975{\natexlab{a}}.

\bibitem[North(1975{\natexlab{b}})]{north1975analytical}
Gerald~R North.
\newblock Analytical solution to a simple climate model with diffusive heat transport.
\newblock \emph{Journal of Atmospheric Sciences}, 32\penalty0 (7):\penalty0 1301--1307, 1975{\natexlab{b}}.

\bibitem[North et~al.(1981)North, Cahalan, and Coakley~Jr]{north1981energy}
Gerald~R North, Robert~F Cahalan, and James~A Coakley~Jr.
\newblock Energy balance climate models.
\newblock \emph{Reviews of Geophysics}, 19\penalty0 (1):\penalty0 91--121, 1981.

\bibitem[Imkeller(2001)]{imkeller2001energy}
Peter Imkeller.
\newblock Energy balance models—viewed from stochastic dynamics.
\newblock In \emph{Stochastic climate models}, pages 213--240. Springer, 2001.

\bibitem[Kalnay(2003)]{kalnay2003atmospheric}
Eugenia Kalnay.
\newblock \emph{Atmospheric modeling, data assimilation and predictability}.
\newblock Cambridge university press, 2003.

\bibitem[Evensen et~al.(2009)]{evensen2009data}
Geir Evensen et~al.
\newblock \emph{Data assimilation: the ensemble Kalman filter}, volume~2.
\newblock Springer, 2009.

\bibitem[Majda and Harlim(2012)]{majda2012filtering}
Andrew~J Majda and John Harlim.
\newblock \emph{Filtering complex turbulent systems}.
\newblock Cambridge University Press, 2012.

\bibitem[Kalman(1960)]{kalman1960new}
Rudolph~Emil Kalman.
\newblock A new approach to linear filtering and prediction problems.
\newblock 1960.

\bibitem[Anderson and Moore(2012)]{anderson2012optimal}
Brian~DO Anderson and John~B Moore.
\newblock \emph{Optimal filtering}.
\newblock Courier Corporation, 2012.

\bibitem[Evensen(1994)]{evensen1994sequential}
Geir Evensen.
\newblock Sequential data assimilation with a nonlinear quasi-geostrophic model using monte carlo methods to forecast error statistics.
\newblock \emph{Journal of Geophysical Research: Oceans}, 99\penalty0 (C5):\penalty0 10143--10162, 1994.

\bibitem[Burgers et~al.(1998)Burgers, Van~Leeuwen, and Evensen]{burgers1998analysis}
Gerrit Burgers, Peter~Jan Van~Leeuwen, and Geir Evensen.
\newblock Analysis scheme in the ensemble kalman filter.
\newblock \emph{Monthly weather review}, 126\penalty0 (6):\penalty0 1719--1724, 1998.

\bibitem[Houtekamer and Mitchell(1998)]{houtekamer1998data}
Peter~L Houtekamer and Herschel~L Mitchell.
\newblock Data assimilation using an ensemble kalman filter technique.
\newblock \emph{Monthly Weather Review}, 126\penalty0 (3):\penalty0 796--811, 1998.

\bibitem[Julier and Uhlmann(1997)]{julier1997new}
Simon~J Julier and Jeffrey~K Uhlmann.
\newblock New extension of the kalman filter to nonlinear systems.
\newblock In \emph{Signal processing, sensor fusion, and target recognition VI}, volume 3068, pages 182--193. Spie, 1997.

\bibitem[Gordon et~al.(1993)Gordon, Salmond, and Smith]{gordon1993novel}
Neil~J Gordon, David~J Salmond, and Adrian~FM Smith.
\newblock Novel approach to nonlinear/non-gaussian bayesian state estimation.
\newblock In \emph{IEE proceedings F (radar and signal processing)}, volume 140, pages 107--113. IET, 1993.

\bibitem[Doucet et~al.(2001)Doucet, De~Freitas, Gordon, et~al.]{doucet2001sequential}
Arnaud Doucet, Nando De~Freitas, Neil~James Gordon, et~al.
\newblock \emph{Sequential Monte Carlo methods in practice}, volume~1.
\newblock Springer, 2001.

\bibitem[Van~Leeuwen(2009)]{van2009particle}
Peter~Jan Van~Leeuwen.
\newblock Particle filtering in geophysical systems.
\newblock \emph{Monthly Weather Review}, 137\penalty0 (12):\penalty0 4089--4114, 2009.

\bibitem[Li et~al.(2015)Li, Bolic, and Djuric]{li2015resampling}
Tiancheng Li, Miodrag Bolic, and Petar~M Djuric.
\newblock Resampling methods for particle filtering: classification, implementation, and strategies.
\newblock \emph{IEEE Signal processing magazine}, 32\penalty0 (3):\penalty0 70--86, 2015.

\bibitem[Bengtsson et~al.(2008)Bengtsson, Bickel, and Li]{bengtsson2008curse}
Thomas Bengtsson, Peter Bickel, and Bo~Li.
\newblock Curse-of-dimensionality revisited: Collapse of the particle filter in very large scale systems.
\newblock In \emph{Probability and statistics: Essays in honor of David A. Freedman}, volume~2, pages 316--335. Institute of Mathematical Statistics, 2008.

\bibitem[Snyder et~al.(2008)Snyder, Bengtsson, Bickel, and Anderson]{snyder2008obstacles}
Chris Snyder, Thomas Bengtsson, Peter Bickel, and Jeff Anderson.
\newblock Obstacles to high-dimensional particle filtering.
\newblock \emph{Monthly Weather Review}, 136\penalty0 (12):\penalty0 4629--4640, 2008.

\bibitem[Pitt and Shephard(1999)]{pitt1999filtering}
Michael~K Pitt and Neil Shephard.
\newblock Filtering via simulation: Auxiliary particle filters.
\newblock \emph{Journal of the American statistical association}, 94\penalty0 (446):\penalty0 590--599, 1999.

\bibitem[de~Freitas et~al.(2000)de~Freitas, Niranjan, Gee, and Doucet]{de2000sequential}
Joao~FG de~Freitas, Mahesan Niranjan, Andrew~H. Gee, and Arnaud Doucet.
\newblock Sequential monte carlo methods to train neural network models.
\newblock \emph{Neural computation}, 12\penalty0 (4):\penalty0 955--993, 2000.

\bibitem[de~Freitas(2003)]{de2003bayesian}
Jo{\~a}o Ferdinando~Gomes de~Freitas.
\newblock \emph{Bayesian methods for neural networks}.
\newblock PhD thesis, University of Cambridge, 2003.

\bibitem[Doucet et~al.(1998)Doucet, Godsill, and Andrieu]{doucet1998sequential}
Arnaud Doucet, Simon~J Godsill, and Christophe Andrieu.
\newblock \emph{On sequential simulation-based methods for Bayesian filtering}.
\newblock Department of Engineering, University of Cambridge Cambridge, UK, 1998.

\bibitem[Van Der~Merwe et~al.(2000)Van Der~Merwe, Doucet, De~Freitas, and Wan]{van2000unscented}
Rudolph Van Der~Merwe, Arnaud Doucet, Nando De~Freitas, and Eric Wan.
\newblock The unscented particle filter.
\newblock \emph{Advances in neural information processing systems}, 13, 2000.

\bibitem[Grewal and Andrews(2010)]{grewal2010applications}
Mohinder~S Grewal and Angus~P Andrews.
\newblock Applications of kalman filtering in aerospace 1960 to the present [historical perspectives].
\newblock \emph{IEEE Control Systems Magazine}, 30\penalty0 (3):\penalty0 69--78, 2010.

\bibitem[Cou{\'e} et~al.(2006)Cou{\'e}, Pradalier, Laugier, Fraichard, and Bessi{\`e}re]{coue2006bayesian}
Christophe Cou{\'e}, C{\'e}dric Pradalier, Christian Laugier, Thierry Fraichard, and Pierre Bessi{\`e}re.
\newblock Bayesian occupancy filtering for multitarget tracking: an automotive application.
\newblock \emph{The International Journal of Robotics Research}, 25\penalty0 (1):\penalty0 19--30, 2006.

\bibitem[Floudas et~al.(2005)Floudas, Polychronopoulos, and Amditis]{floudas2005survey}
Nikolaos Floudas, Aris Polychronopoulos, and Angelos Amditis.
\newblock A survey of filtering techniques for vehicle tracking by radar equipped automotive platforms.
\newblock In \emph{2005 7th International Conference on Information Fusion}, volume~2, pages 8--pp. IEEE, 2005.

\bibitem[Gy{\"o}rgy et~al.(2014)Gy{\"o}rgy, Kelemen, and D{\'a}vid]{gyorgy2014unscented}
Katalin Gy{\"o}rgy, Andr{\'a}s Kelemen, and L{\'a}szl{\'o} D{\'a}vid.
\newblock Unscented kalman filters and particle filter methods for nonlinear state estimation.
\newblock \emph{Procedia Technology}, 12:\penalty0 65--74, 2014.

\bibitem[Roth et~al.(2017)Roth, Hendeby, Fritsche, and Gustafsson]{roth2017ensemble}
Michael Roth, Gustaf Hendeby, Carsten Fritsche, and Fredrik Gustafsson.
\newblock The ensemble kalman filter: a signal processing perspective.
\newblock \emph{EURASIP Journal on Advances in Signal Processing}, 2017:\penalty0 1--16, 2017.

\bibitem[Chatzi and Smyth(2009)]{chatzi2009unscented}
Eleni~N Chatzi and Andrew~W Smyth.
\newblock The unscented kalman filter and particle filter methods for nonlinear structural system identification with non-collocated heterogeneous sensing.
\newblock \emph{Structural Control and Health Monitoring: The Official Journal of the International Association for Structural Control and Monitoring and of the European Association for the Control of Structures}, 16\penalty0 (1):\penalty0 99--123, 2009.

\bibitem[Evensen and Van~Leeuwen(1996)]{evensen1996assimilation}
Geir Evensen and Peter~Jan Van~Leeuwen.
\newblock Assimilation of geosat altimeter data for the agulhas current using the ensemble kalman filter with a quasigeostrophic model.
\newblock \emph{Monthly Weather Review}, 124\penalty0 (1):\penalty0 85--96, 1996.

\bibitem[Sebacher et~al.(2013)Sebacher, Hanea, and Heemink]{sebacher2013probabilistic}
Bogdan Sebacher, Remus Hanea, and Arnold Heemink.
\newblock A probabilistic parametrization for geological uncertainty estimation using the ensemble kalman filter (enkf).
\newblock \emph{Computational Geosciences}, 17:\penalty0 813--832, 2013.

\bibitem[Szunyogh et~al.(2005)Szunyogh, Kostelich, Gyarmati, Patil, Hunt, Kalnay, Ott, and Yorke]{szunyogh2005assessing}
Istvan Szunyogh, Eric~J Kostelich, G~Gyarmati, DJ~Patil, Brian~R Hunt, Eugenia Kalnay, Edward Ott, and James~A Yorke.
\newblock Assessing a local ensemble kalman filter: Perfect model experiments with the national centers for environmental prediction global model.
\newblock \emph{Tellus A: Dynamic Meteorology and Oceanography}, 57\penalty0 (4):\penalty0 528--545, 2005.

\bibitem[Aanonsen et~al.(2009)Aanonsen, N{\oe}vdal, Oliver, Reynolds, and Vall{\`e}s]{aanonsen2009ensemble}
Sigurd~I Aanonsen, Geir N{\oe}vdal, Dean~S Oliver, Albert~C Reynolds, and Brice Vall{\`e}s.
\newblock The ensemble kalman filter in reservoir engineering—a review.
\newblock \emph{Spe Journal}, 14\penalty0 (03):\penalty0 393--412, 2009.

\bibitem[Houtekamer et~al.(2005)Houtekamer, Mitchell, Pellerin, Buehner, Charron, Spacek, and Hansen]{houtekamer2005atmospheric}
Peter~L Houtekamer, Herschel~L Mitchell, G{\'e}rard Pellerin, Mark Buehner, Martin Charron, Lubos Spacek, and Bjarne Hansen.
\newblock Atmospheric data assimilation with an ensemble kalman filter: Results with real observations.
\newblock \emph{Monthly weather review}, 133\penalty0 (3):\penalty0 604--620, 2005.

\bibitem[Whitaker et~al.(2008)Whitaker, Hamill, Wei, Song, and Toth]{whitaker2008ensemble}
Jeffrey~S Whitaker, Thomas~M Hamill, Xue Wei, Yucheng Song, and Zoltan Toth.
\newblock Ensemble data assimilation with the ncep global forecast system.
\newblock \emph{Monthly Weather Review}, 136\penalty0 (2):\penalty0 463--482, 2008.

\bibitem[Szunyogh et~al.(2008)Szunyogh, Kostelich, Gyarmati, Kalnay, Hunt, Ott, Satterfield, and Yorke]{szunyogh2008local}
Istvan Szunyogh, Eric~J Kostelich, Gyorgyi Gyarmati, Eugenia Kalnay, Brian~R Hunt, Edward Ott, Elizabeth Satterfield, and James~A Yorke.
\newblock A local ensemble transform kalman filter data assimilation system for the ncep global model.
\newblock \emph{Tellus A: Dynamic Meteorology and Oceanography}, 60\penalty0 (1):\penalty0 113--130, 2008.

\bibitem[Houtekamer et~al.(2009)Houtekamer, Mitchell, and Deng]{houtekamer2009model}
PL~Houtekamer, Herschel~L Mitchell, and Xingxiu Deng.
\newblock Model error representation in an operational ensemble kalman filter.
\newblock \emph{Monthly Weather Review}, 137\penalty0 (7):\penalty0 2126--2143, 2009.

\bibitem[Thrun(2002)]{thrun2002particle}
Sebastian Thrun.
\newblock Particle filters in robotics.
\newblock In \emph{UAI}, volume~2, pages 511--518. Citeseer, 2002.

\bibitem[Fox et~al.(2001)Fox, Thrun, Burgard, and Dellaert]{fox2001particle}
Dieter Fox, Sebastian Thrun, Wolfram Burgard, and Frank Dellaert.
\newblock Particle filters for mobile robot localization.
\newblock In \emph{Sequential Monte Carlo methods in practice}, pages 401--428. Springer, 2001.

\bibitem[Kearns(2005)]{Kearns2005}
Michael Kearns.
\newblock Filtering in finance.
\newblock \url{https://www.cis.upenn.edu/~mkearns/finread/filtering_in_finance.pdf}, 2005.
\newblock Accessed: 2023-04-10.

\bibitem[Javaheri et~al.(2003)Javaheri, Lautier, and Galli]{javaheri2003filtering}
Alireza Javaheri, Delphine Lautier, and Alain Galli.
\newblock Filtering in finance.
\newblock \emph{Wilmott}, 3:\penalty0 67--83, 2003.

\bibitem[Wells(2013)]{wells2013kalman}
Curt Wells.
\newblock \emph{The Kalman filter in finance}, volume~32.
\newblock Springer Science \& Business Media, 2013.

\bibitem[Date and Ponomareva(2011)]{date2011linear}
Paresh Date and Ksenia Ponomareva.
\newblock Linear and non-linear filtering in mathematical finance: a review.
\newblock \emph{IMA Journal of Management Mathematics}, 22\penalty0 (3):\penalty0 195--211, 2011.

\bibitem[Zhan et~al.(2008)Zhan, Xin, and Jianwei]{zhan2008modified}
Ronohui Zhan, Qin Xin, and Wan Jianwei.
\newblock Modified unscented particle filter for nonlinear bayesian tracking.
\newblock \emph{Journal of Systems Engineering and Electronics}, 19\penalty0 (1):\penalty0 7--14, 2008.

\bibitem[Havangi(2018)]{havangi2018target}
R~Havangi.
\newblock Target tracking based on improved unscented particle filter with markov chain monte carlo.
\newblock \emph{IETE Journal of Research}, 64\penalty0 (6):\penalty0 873--885, 2018.

\bibitem[Zhang et~al.(2018)Zhang, Miao, Zhang, and Liu]{zhang2018improved}
Heng Zhang, Qiang Miao, Xin Zhang, and Zhiwen Liu.
\newblock An improved unscented particle filter approach for lithium-ion battery remaining useful life prediction.
\newblock \emph{Microelectronics Reliability}, 81:\penalty0 288--298, 2018.

\bibitem[Miao et~al.(2013)Miao, Xie, Cui, Liang, and Pecht]{miao2013remaining}
Qiang Miao, Lei Xie, Hengjuan Cui, Wei Liang, and Michael Pecht.
\newblock Remaining useful life prediction of lithium-ion battery with unscented particle filter technique.
\newblock \emph{Microelectronics Reliability}, 53\penalty0 (6):\penalty0 805--810, 2013.

\bibitem[Wang and Chen(2020)]{wang2020framework}
Yujie Wang and Zonghai Chen.
\newblock A framework for state-of-charge and remaining discharge time prediction using unscented particle filter.
\newblock \emph{Applied Energy}, 260:\penalty0 114324, 2020.

\bibitem[MIT(2023)]{MITClimateModelFall2023}
MIT.
\newblock Our first climate model - computational thinking.
\newblock \url{https://computationalthinking.mit.edu/Fall23/climate_science/our_first_climate_model}, 2023.
\newblock Accessed: 2024-03-22.

\bibitem[Sherwood et~al.(2020)Sherwood, Webb, Annan, Armour, Forster, Hargreaves, Hegerl, Klein, Marvel, Rohling, et~al.]{sherwood2020assessment}
Steven~C Sherwood, Mark~J Webb, James~D Annan, Kyle~C Armour, Piers~M Forster, Julia~C Hargreaves, Gabriele Hegerl, Stephen~A Klein, Kate~D Marvel, Eelco~J Rohling, et~al.
\newblock An assessment of earth's climate sensitivity using multiple lines of evidence.
\newblock \emph{Reviews of Geophysics}, 58\penalty0 (4):\penalty0 e2019RG000678, 2020.

\bibitem[Rahmstorf(2007)]{rahmstorf2007}
Stefan Rahmstorf.
\newblock A semi-empirical approach to projecting future sea-level rise.
\newblock \emph{Science}, 315\penalty0 (5810):\penalty0 368--370, 2007.
\newblock \doi{10.1126/science.1135456}.

\bibitem[Church and White(2011)]{church2011sea}
John~A Church and Neil~J White.
\newblock Sea-level rise from the late 19th to the early 21st century.
\newblock \emph{Surveys in geophysics}, 32:\penalty0 585--602, 2011.

\bibitem[Church et~al.(2001)Church, Gregory, Huybrechts, Kuhn, Lambeck, Nhuan, Qin, and Woodworth]{church2001changes}
John~A Church, Jonathan~M Gregory, Philippe Huybrechts, Michael Kuhn, Kurt Lambeck, Mai~T Nhuan, Dahe Qin, and Phil~L Woodworth.
\newblock Changes in sea level.
\newblock In \emph{, in: JT Houghton, Y. Ding, DJ Griggs, M. Noguer, PJ Van der Linden, X. Dai, K. Maskell, and CA Johnson (eds.): Climate Change 2001: The Scientific Basis: Contribution of Working Group I to the Third Assessment Report of the Intergovernmental Panel}, pages 639--694. 2001.

\bibitem[Aral et~al.(2012)Aral, Guan, and Chang]{aral2012dynamic}
Mustafa~M Aral, Jiabao Guan, and Biao Chang.
\newblock Dynamic system model to predict global sea-level rise and temperature change.
\newblock \emph{Journal of Hydrologic Engineering}, 17\penalty0 (2):\penalty0 237--242, 2012.

\bibitem[Wan and Van Der~Merwe(2000)]{wan2000unscented}
Eric~A Wan and Rudolph Van Der~Merwe.
\newblock The unscented kalman filter for nonlinear estimation.
\newblock In \emph{Proceedings of the IEEE 2000 adaptive systems for signal processing, communications, and control symposium (Cat. No. 00EX373)}, pages 153--158. Ieee, 2000.

\bibitem[Kitagawa(1996)]{kitagawa1996monte}
Genshiro Kitagawa.
\newblock Monte carlo filter and smoother for non-gaussian nonlinear state space models.
\newblock \emph{Journal of computational and graphical statistics}, 5\penalty0 (1):\penalty0 1--25, 1996.

\bibitem[Carpenter et~al.(1999)Carpenter, Clifford, and Fearnhead]{carpenter1999improved}
James Carpenter, Peter Clifford, and Paul Fearnhead.
\newblock Improved particle filter for nonlinear problems.
\newblock \emph{IEE Proceedings-Radar, Sonar and Navigation}, 146\penalty0 (1):\penalty0 2--7, 1999.

\bibitem[Bitzer(2022)]{bitzer_ukf}
Sebastian Bitzer.
\newblock Ukf exposed.
\newblock GitHub repository, 2022.
\newblock URL \url{https://nbviewer.org/github/sbitzer/UKF-exposed/blob/master/UKF.ipynb}.

\bibitem[Gillijns et~al.(2006)Gillijns, Mendoza, Chandrasekar, De~Moor, Bernstein, and Ridley]{gillijns2006ensemble}
Steven Gillijns, O~Barrero Mendoza, Jaganath Chandrasekar, BLR De~Moor, Dennis~S Bernstein, and A~Ridley.
\newblock What is the ensemble kalman filter and how well does it work?
\newblock In \emph{2006 American control conference}, pages 6--pp. IEEE, 2006.

\end{thebibliography}

\end{document}